\documentclass{rfpt}

\usepackage{graphicx}

\usepackage[latin1]{inputenc}   

\usepackage{amssymb,amsthm}

\usepackage{amsmath}
\usepackage{pgfplots}
\usepackage{subcaption}
\usepackage{color}
\usepackage{tikz}
\usetikzlibrary{arrows,patterns}
\usepackage{rotating}
\usepackage{verbatim}
\usepackage{url}
\usepackage{float}
\usepackage{array}
\usepackage[normalem]{ulem}
\usepackage{ragged2e}

\pgfplotsset{compat=1.8}
\usepgfplotslibrary{statistics}

\usepackage{array}
\newcolumntype{L}[1]{>{\raggedright\let\newline\\\arraybackslash\hspace{0pt}}m{#1}}
\newcolumntype{C}[1]{>{\centering\let\newline\\\arraybackslash\hspace{0pt}}m{#1}}
\newcolumntype{R}[1]{>{\raggedleft\let\newline\\\arraybackslash\hspace{0pt}}m{#1}}
\newcolumntype{M}{>{\centering\arraybackslash} m{3cm} } 

\begin{document}

\begin{frontmatter}



\title{Weighted simplicial complex reconstruction from mobile laser scanning using sensor topology}

\author[A1]{StÈphane Guinard}
\author[A1]{Bruno Vallet}

\address[A1]{Universit\'e Paris-Est, LASTIG MATIS, IGN, ENSG, \\
	73 avenue de Paris, 94160 Saint-Mand\'e, France}

\begin{resume}
Nous prÈsentons une nouvelle mÈthode pour la reconstruction de complexes simpliciaux (ensembles de points, segments et triangles) ‡ partir de nuages de points 3D obtenus par LiDAR mobile. Notre mÈthode utilise la topologie inhÈrente au capteur LiDAR pour dÈfinir une relation spatiale entre les points. Pour cela, nous examinons chaque connexion possible entre points, pondÈrÈe en fonction de sa distance au capteur, et les filtrons en privilÈgiant les structures collinÈaires, ou perpendiculaires aux impulsions du laser. Ensuite, nous crÈons et filtrons des triangles pour chaque triplet de segments connectÈs entre eux, en fonction de leur coplanaritÈ locale. Nous comparons nos rÈsultats ‡ une reconstruction non pondÈrÈe d'un complexe simplicial.
\end{resume}

\begin{motscle}
Complexes simpliciaux \sep reconstruction 3D \sep nuages de points \sep Lidar mobile \sep topologie capteur
\end{motscle}

\begin{abstract}
We propose a new method for the reconstruction of simplicial complexes (combining points, edges and triangles) from 3D point clouds from Mobile Laser Scanning (MLS). Our method uses the inherent topology of the MLS sensor to define a spatial adjacency relationship between points. We then investigate each possible connexion between adjacent points, weighted according to its distance to the sensor, and filter them by searching collinear structures in the scene, or structures perpendicular to the laser beams. Next, we create and filter triangles for each triplet of self-connected edges and according to their local planarity.  We compare our results to an unweighted simplicial complex reconstruction.\end{abstract}


\end{frontmatter}

\section{Introduction}

LiDAR scanning technologies have become a widespread and direct mean for acquiring a precise sampling of the geometry of scenes of interest. However, unlike images, LiDAR point clouds do not always have a natural topology (4- or 8-neighborhoods for images) allowing to recover the continuous nature of the acquired scenes from the individual samples. This is why a large amount of research work has been dedicated into recovering a continuous surface from a cloud of point samples, which is a central problem in geometry processing. Surface reconstruction generally aims at reconstructing triangulated surface meshes from point clouds, as they are the most common numerical representation for surfaces in 3D, thus well adapted for further processing. The reconstruction of surface meshes has numerous applications in various domains:
\begin{itemize}
\item Visualization: a surface mesh is much more adapted to visualization than a point cloud, as the visible surface is interpolated between points, allowing for a continuous representation of the real surface, and enabling the appraisal of occlusions, thus to render only the visible parts of the scene.
\item Estimation of differential quantities such as surface normals and curvatures.
\item Texturing: a surface mesh can be textured (by applying images on it) allowing for photo-realistic rendering. In particular, when multiple images of the acquired scene exists, texturing allows to fusion and blend them all into a single 3D representation.
\item Shape and object detection and reconstruction: these high level processes benefit from surface reconstruction since it solves the basic geometric ambiguity (which points are connected by a real surface in the real scene ? ).
\end{itemize}

\begin{figure*}[t]
\centering
\includegraphics[width=1\textwidth]{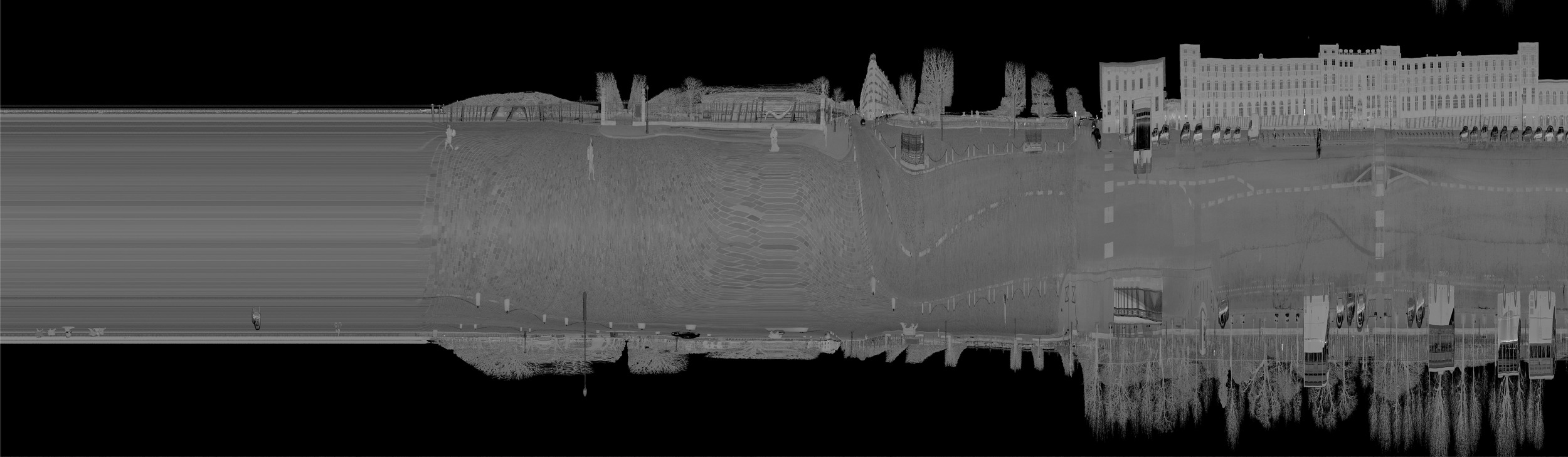}
\caption{Echo intensity of an MLS displayed in sensor topology: vertical axis is the angle $\theta$, horizontal axis is the line number, equivalent to time as the scanner acquires a constant number of lines per second. Horizontal resolution depends on vehicle speed (the left part is constant because the vehicle is stopped).}
\label{fig:sensorTopo}
\end{figure*}

\begin{figure*}[t]
\centering
\includegraphics[width=1\textwidth]{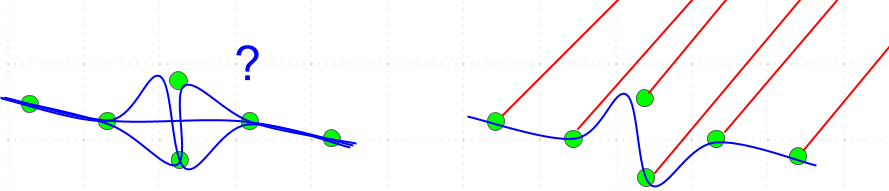}
\caption{Left: a 2D point cloud (green) and possible reconstructions (blue). Right: knowing the LiDAR rays allows to solve the ambiguity}
\label{fig:ambig}
\end{figure*}

In practice, existing surface reconstruction algorithms often consider that their input is a set of $(x,y,z)$ coordinates, possibly with normals. However, most LiDAR scanning technologies provide more than that: the sensors have a logic of acquisition that provides a sensor topology \citep{xiao2013change,vallet2015terramobilita}. For instance, planar scanners acquire points along a line that advances with the platform (plane, car, ...) it is mounted on. Thus each point can be naturally connected to the one before and after him along the line, and to its equivalent in the previous and next lines (see Figure \ref{fig:sensorTopo}).
Fixed LiDARs scan in spherical $(\theta, \phi)$ coordinates which also imply a natural connection of each point to the previous and next along these two angles. Some scanner manufacturers exploit this topology by proposing visualization and processing tools in 2.5D (depth images in $(\theta, \phi)$) rather than 3D. Moreover, LiDAR scanning can provide a meaningful information that is the position of the LiDAR sensor for each point, resulting in a ray along which we are sure that space is empty. This information can also disambiguate surface reconstruction as illustrated in Figure \ref{fig:ambig}. This is why we decided to investigate the use of the sensor topology inherent to an MLS, to perform a 3D reconstruction of a point cloud.

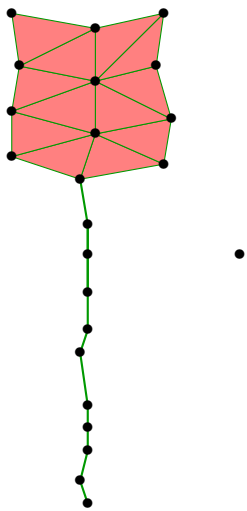
\begin{figure}[t]
\centering
\begin{tikzpicture}


\draw[thick,color=green!60!black] (0,2) -- (-.1,1.7);
\draw[thick,color=green!60!black] (0,1) -- (-.1,1.7);
\draw[thick,color=green!60!black] (0,1) -- (0,.7);
\draw[thick,color=green!60!black] (0,.4) -- (0,.7);
\draw[thick,color=green!60!black] (0,.4) -- (-0.1,0);
\draw[thick,color=green!60!black] (0,-.3) -- (-0.1,0);
\draw[thick,color=green!60!black] (0,3) -- (0,2.5);
\draw[thick,color=green!60!black] (0,3) -- (0,3.4);
\draw[thick,color=green!60!black] (0,3) -- (0,3.4);
\draw[thick,color=green!60!black] (0,3) -- (0,2);
\draw[thick,color=green!60!black] (-.1,4) -- (0,3.4);

\draw[fill=red!50,draw=green!60!black] (0.1,4.6) -- (-0.1,4) -- (1,4.2) -- cycle;
\draw[fill=red!50,draw=green!60!black] (1.1,4.8) -- (0.1,4.6) -- (1,4.2) -- cycle;
\draw[fill=red!50,draw=green!60!black] (1.1,4.8) -- (0.1,4.6) -- (0.1,5.3) -- cycle;
\draw[fill=red!50,draw=green!60!black] (1.1,4.8) -- (0.9,5.5) -- (0.1,5.3) -- cycle;
\draw[fill=red!50,draw=green!60!black] (.9,5.5) -- (1,6.2) -- (0.1,5.3) -- cycle;
\draw[fill=red!50,draw=green!60!black] (0.1,5.3) -- (1,6.2) -- (0.1,6) -- cycle;

\draw[fill=red!50,draw=green!60!black] (0.1,4.6) -- (-0.1,4) -- (-1,4.3) -- cycle;
\draw[fill=red!50,draw=green!60!black] (0.1,4.6) -- (-1,4.9) -- (-1,4.3) -- cycle;
\draw[fill=red!50,draw=green!60!black] (-1,4.9) -- (0.1,5.3) -- (0.1,4.6) -- cycle;
\draw[fill=red!50,draw=green!60!black] (-.9,5.5) -- (0.1,5.3) -- (-1,4.9) -- cycle;
\draw[fill=red!50,draw=green!60!black] (-.9,5.5) -- (0.1,5.3) -- (.1,6) -- cycle;
\draw[fill=red!50,draw=green!60!black] (-1,6.2) -- (-.9,5.5) -- (.1,6) -- cycle;

\node at (0.1,5.3) {$\bullet$};
\node at (0.1,4.6) {$\bullet$};
\node at (0.1,6) {$\bullet$};
\node at (-0.1,4) {$\bullet$};
\node at (0,3) {$\bullet$};
\node at (0,2) {$\bullet$};
\node at (0,1) {$\bullet$};
\node at (0,3.4) {$\bullet$};
\node at (0,2.5) {$\bullet$};
\node at (-0.1,1.7) {$\bullet$};
\node at (1,4.2) {$\bullet$};

\node at (2,3) {$\bullet$};
\node at (0,-.3) {$\bullet$};
\node at (-0.1,0) {$\bullet$};
\node at (0,0.7) {$\bullet$};
\node at (0,0.4) {$\bullet$};
\node at (-1,4.3) {$\bullet$};
\node at (1.1,4.8) {$\bullet$};
\node at (0.9,5.5) {$\bullet$};
\node at (-1,4.9) {$\bullet$};
\node at (-0.9,5.5) {$\bullet$};
\node at (1,6.2) {$\bullet$};
\node at (-1,6.2) {$\bullet$};

\end{tikzpicture}

\caption{A simplicial complex consists of simplices of dimension 0 (points, black), 1 (edges, green) and 2 (triangles, red).}
\label{fig:simplicialComplexe}
\end{figure}

Secondly, the geometry processing community has mainly focused on reconstruction of rather smooth objects, possibly with sharp edges, but with a sampling density sufficient to consider that the object is a 2-manifold, which means that it is locally 2-dimensional. Thus these methods do not extend well to real scenes where such a guarantee is hardly possible. In particular, scans including poles, power lines, wires, ... almost never allow to create triangles on these structures because their widths (a few mm to a few cm) is much smaller than the scanning resolution. Scans of highly detailed structures (such as tree foliage for instance) even have a 0-dimensional nature: individual points should not even be connected to any of their neighbors. Applying the Nyquist-Shannon theorem to the range in sensor space tells us that if the geometric frequency (frequency of the range signal in sensor space) is higher than half the sampling frequency (frequency of the samples in sensor space), then some (geometric) signal will be lost. This happens in the cases stated above for instance. Because of this, we should aim at reconstructing triangles only when the Shannon condition is met in the two dimensions, but only edges when the geometric frequency is too high in 1 dimension and points when the geometric frequency is too high in the 2 dimensions. Triangles, edges and points are called simplices, which are characterized by their dimension $d$ ($0=$ vertices, $1=$ edges, $2=$ triangles). If we add the constraint that edges can only meet at a vertex and triangles can only meet at an edge or vertex, the resulting mathematical object is called a \textit{simplicial complex} as illustrated in Figure \ref{fig:simplicialComplexe}. The aim of this paper is to propose a method to reconstruct such simplicial complexes from a LiDAR scan.

\section{State of the art}

3D surface mesh reconstruction from point clouds has been a major issue in geometry processing for the last decades.
3D reconstruction can be performed from oriented \citep{kazhdan2013screened} or unoriented point sets \citep{alliez2007voronoi}. Data itself can come from various sources: Terrestrial Laser Scanning \citep{pu2009knowledge}, Aerial Laser Scanning \citep{dorninger2008comprehensive, zhu2014use} or Mobile Laser Scanning. We refer the reader to  \citet{berger2014state} for a general review of surface reconstruction methodologies and focus our state of the art on surface reconstruction from Mobile Laser Scanning (MLS) and on simplicial complexes reconstruction, which are the two specificities of our approach.

MLS have been used for the past years mostly for the modeling of outdoors environments, usually in urban scenes. An automatically-generated grammar for the reconstruction of buildings has been proposed by \citet{becker2009grammar}, whereas \citet{morsdorf2004lidar} and \citet{rutzinger2011tree} focus more specifically on tree shapes reconstruction. MLS has also been useful for specific indoor environments: \citet{zlot2014efficient} used an MLS in an underground mine to obtain a 3D model of the tunnels.

The utility of simplicial complexes for the reconstruction of 3D point clouds has been expressed by \citet{popovic1997progressive} as a generalization manner to simplify 3D meshes. Simplicial complexes are also used to simplify defect-laden point sets as a way to be robust to noise and outliers using optimal transport \citep{de2011optimal,digne2014feature} or alpha-shapes \citep{bernardini1997sampling}. 

A first approach has been presented in \citet{guinard2018sensor}, but suffers from its lack of global knowledge. Our main goal is to produce a 3D reconstruction of a scene using the sensor topology of an MLS to build a simplicial complex, which should be independent from the sampling.
Our goal is to improve this method so that the reconstruction criteria is adaptive to the structure of the cloud. The aim of this paper is to propose a reconstruction method combining the following advantages:
\begin{enumerate}
\item Reconstruction of a simplicial complex instead of a surface mesh, adapting the local dimension to that of the local structure.
\item Exploiting the sensor topology both to solve ambiguities and to speed up computations.
\item Exploiting the global structure of the cloud.
\end{enumerate}

\section{Methodology}

As explained above, sensor topology yields in general a regular mesh structure with a 6-neighborhood that can be used to perform a surface mesh reconstruction. This reconstruction is however very poor as all depth discontinuities will be meshed, hence very elongated triangles will be created between objects and their background. This section investigates criteria to remove these specific triangles, while possibly keeping some of their edges. As all input points will be kept, the resulting reconstruction combines points and edges and triangles based on these points, which is called a \textit{simplicial complex} in mathematics.

\subsection{Objectives}



\begin{figure}[t]
\centering
\begin{subfigure}[t]{.15\textwidth}
\centering
  \begin{tikzpicture}[scale=0.2]
    \draw[-latex,black] (0,5) -- (0,2);
    \draw[-latex,black] (2,5) -- (2,1.6);
    \draw[-latex,black] (4,5) -- (4,0.6);
    \draw[-latex,black] (6,5) -- (6,-3.2);
    \draw[-latex,black] (8,5) -- (8,-2.3);
    \draw (-1,2) ..controls +(5,0.3) and +(5.5,0.3).. (2,-2); 
    \draw (11,-1.5) ..controls (6,-2.5) and +(5,0.5).. (1,-4.5); 
  \end{tikzpicture}
  \caption{Separation case}
  \label{fig:sep}
  \end{subfigure}%
  \begin{subfigure}[t]{.15\textwidth} 
  \centering
  \begin{tikzpicture}[scale=0.2]
    \draw[-latex,black] (0,6) -- (0,2);
    \draw[-latex,black] (2,6) -- (2,1.6);
    \draw[-latex,black] (4,6) -- (4,0.6);
    \draw[-latex,black] (6,6) -- (6,-3.2);
    \draw[-latex,black] (8,6) -- (8,-2.3);
    \draw plot [smooth] coordinates {(-1,2) (0,2) (2,1.6) (4,.6) (5,-.5) (6,-3.2) (8,-2.3) (11,-1)};
  \end{tikzpicture}
  \caption{Ambiguous case}
  \label{fig:ambigu}
  \end{subfigure}%
  \begin{subfigure}[t]{0.15\textwidth}
  \centering
  \begin{tikzpicture}[scale=0.2]
    \draw[-latex,black] (10,4) -- (10,3);
    \draw[-latex,black] (2,4) -- (2,-5.2);
    \draw[-latex,black] (4,4) -- (4,-4.8);
    \draw[-latex,black] (6,4) -- (6,-1);
    \draw[-latex,black] (8,4) -- (8,2.6);
    \draw plot [smooth] coordinates {(1,-5.2) (2,-5.2) (4,-4.8) (6,-1) (8,2.6) (10,3) (11,3)}; 
  \end{tikzpicture}
  \caption{Non-\\ separation case}
  \label{fig:no-sep}
  \end{subfigure}
\caption{Illustration of the two cases of important depth difference. The arrows represent the laser beams. Figures \ref{fig:sep} and \ref{fig:ambigu} show the cases where two neighboring echoes have a huge depth difference. They can either fall on two different objects or on a same object and we have no hint to distinguish these two cases. 
Figure \ref{fig:no-sep} shows the case where three or more echoes are approximately aligned, with a huge depth difference. In this case we want to reconstruct edges between these echoes because it may correspond to a grazing surface.}
\label{fig:sep-cases}
\end{figure}
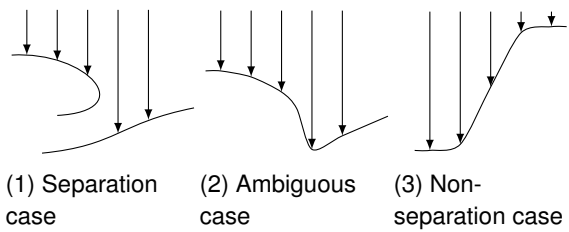

Our main objective is to determine which adjacent points (in sensor topology) should be connected to form edges and triangles. We consider that we may be facing a discontinuity when the depth difference between two neighboring echoes is high. This depth difference is computed from a sensor viewpoint, which implies that a large depth difference may correspond to two cases: either the echoes fell on two different objects with a notable depth difference
, or they fell on a grazing surface (nearly parallel to the laser beams direction) as shown in figure \ref{fig:sep-cases}. When two neighboring echoes have a large depth difference, there is no way we can guess whether they are located on two separate objects (\ref{fig:sep}) or a grazing surface (\ref{fig:ambigu}). The core idea of our filtering is that the only hint we can rely on to distinguish between these two cases is that if at least three echoes with a large depth differences are aligned (\ref{fig:no-sep}), we probably are in the grazing surface case rather than on separate objects.

\begin{figure*}[t]
\centering
\begin{subfigure}{0.48\textwidth}
\includegraphics[scale=.18]{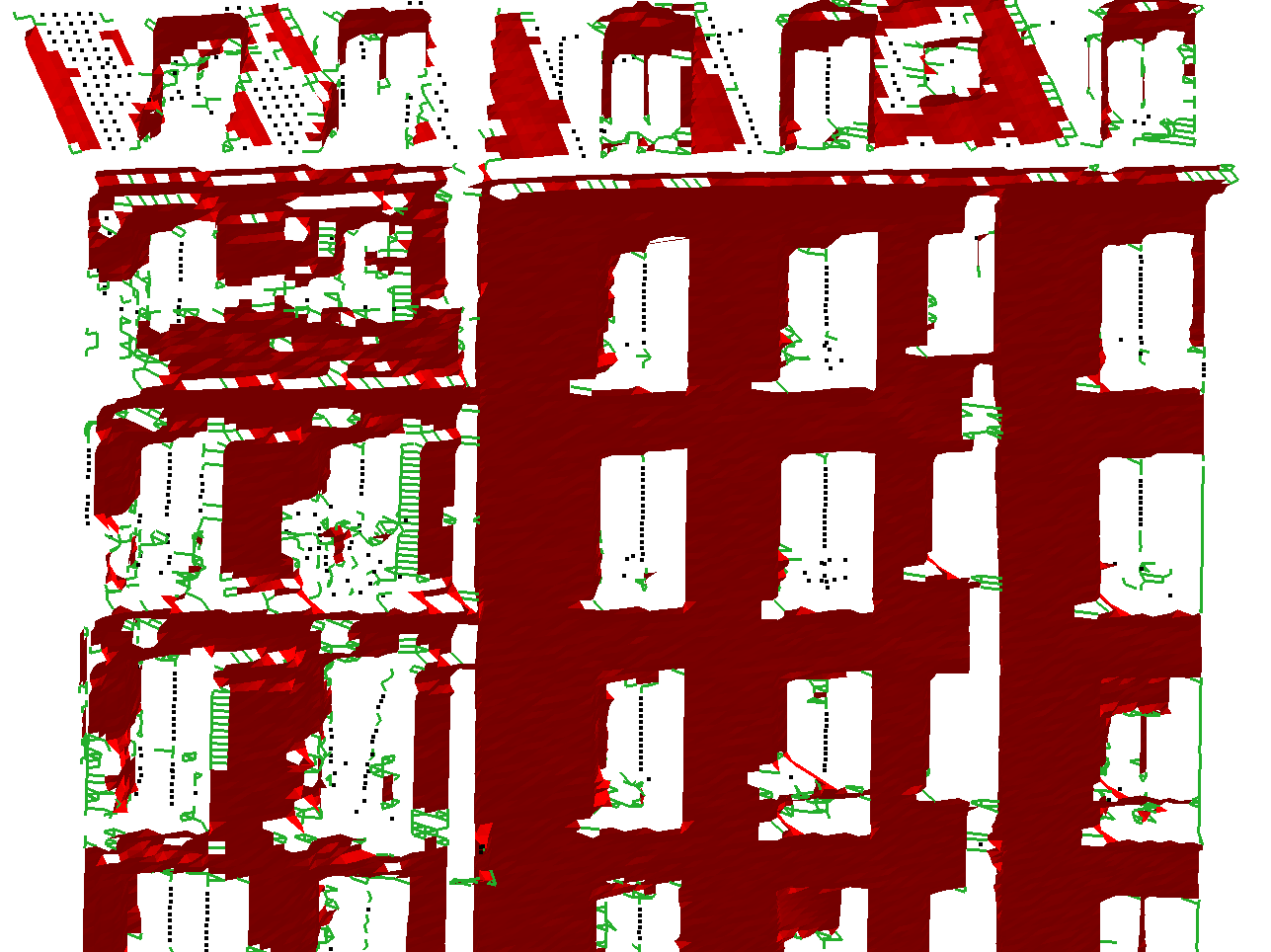}
\caption{Unweighted reconstruction.}
\end{subfigure}
\begin{subfigure}{0.48\textwidth}
\includegraphics[scale=.18]{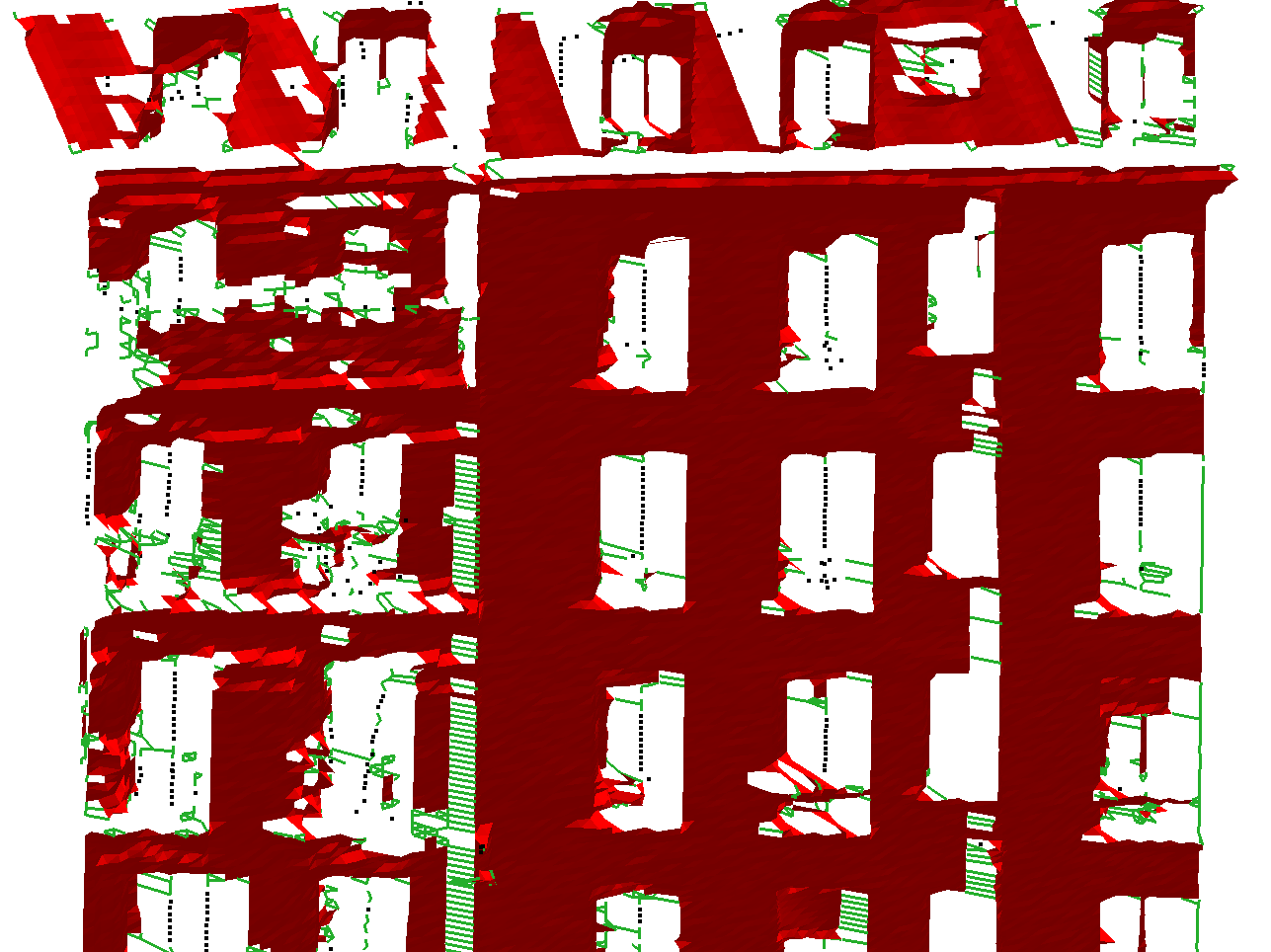}
\caption{Weighted reconstruction}
\end{subfigure}
\caption{Utility of the weighted simplicial complex reconstruction in areas far from the laser. Here the scene represents a facade in the grazing surface case. The triangles are shown in red, the edges in green and the points in black.}
\label{fig:weight}
\end{figure*}

Unlike \citet{guinard2018sensor}, whose algorithm was independent from the sampling, we make the assumption that the distance of some part of the scene to the laser, and the regularity of the emissions of the laser lead to points being naturally further one of the other in some areas far from the sensor. The influence of this phenomenon is especially visible on figure \ref{fig:weight}, which represent our reconstructions on such a part of a scene. The left part is computed without taking the sampling variation into account. We clearly see that the algorithm misses a large part of the roof which we want to retrieve thanks to a weighting of edges. Moreover, the reconstruction on the left is very holed, whereas the one on the right shows a greater regularity. Thus, the criteria for the filtering of possible connexions in \citep{guinard2018sensor} has to be modified in order to take into account this depth difference varying according to the distance to the laser. We precise that from \citep{guinard2018sensor}, only the so-called $C_0$ regularity is modified. The detail of the algorithm is presented in section \ref{sec:rec}.

To perform our reconstruction, we consider each echo as an independent point. First, we define a neighborhood relationship between echoes in the sensor topology. After that, we create edges, based on the echoes and weighted according to their distance to the sensor. Last, we add triangles based on the edges.

\captionsetup[figure]{
justification=raggedright
}
\captionsetup[figure]{
justification=justified
}

\subsection{Neighborhood in sensor topology}

The sensors used to capture point clouds often have an inherent topology. Mobile Laser Scanners sample a regular grid in ($\theta, t$) where $\theta$ is the rotation angle of the laser beam and $t$ the instant of acquisition. Because the vehicle moves at a varying speed (to adapt to the traffic and respect the circulation rules) and may rotate,  the sampling is however not uniform in space.
In general, the number $N_{p}$ of pulses for a 2$\pi$ rotation in $\theta$ is not an integer so a pulse $P_i$ has six neighbors $P_{i-1}$, $P_{i+1}$, $P_{i-n}$, $P_{i-n-1}$, $P_{i+n}$, $P_{i+n+1}$ where $n=\lfloor N_p \rfloor$ is the integer part of $N_p$ as illustrated on figure \ref{fig:6v}.

However, this topology concerns emitted pulses, not recorded echoes. One pulse might have 0 echo (no target hit) or up to 8 as most modern scanners can record multiple echoes for one pulse if the laser beam intersected several targets, which is very frequent in the vegetation or transparent objects for instance.

\begin{figure}[t]
\centering
\begin{subfigure}[t]{0.225\textwidth}
\centering
\resizebox{0.95\textwidth}{0.95\textwidth}{
\begin{tikzpicture}
\draw[-latex] (0,0) -- (0,3.2);
\draw[-latex] (0,0) -- (5.2,0);
\node at (-0.2,1.5) {$\theta$};
\node at (2.5,-0.2) {$\mathit{t}$};

\node at (0.2,0) {$\bullet$};
\node at (0.4,0.8) {$\bullet$};
\node at (0.6,1.6) {$\bullet$};
\node at (0.8,2.4) {$\bullet$};
\node at (1.8,0.4) {$\bullet$};
\node at (2,1.2) {\textcolor{red!60!black}{$\bullet$}};
\node at (2.2,2) {$\bullet$};
\node at (2.4,2.8) {$\bullet$};
\node at (3.4,0) {$\bullet$};
\node at (3.6,0.8) {$\bullet$};
\node at (3.8,1.6) {$\bullet$};
\node at (4,2.4) {$\bullet$};

\draw[-latex,red!60!black] (2,1.2) -- (0.4,0.8);
\draw[-latex,red!60!black] (2,1.2) -- (0.6,1.6);
\draw[-latex,red!60!black] (2,1.2) -- (1.8,0.4);
\draw[-latex,red!60!black] (2,1.2) -- (2.2,2);
\draw[-latex,red!60!black] (2,1.2) -- (3.6,0.8);
\draw[-latex,red!60!black] (2,1.2) -- (3.8,1.6);
\end{tikzpicture}}
\caption{The pulse sensor topology forms a 6-neighborhood}
\label{fig:6v}
\end{subfigure}%
\hspace{.1cm}
\begin{subfigure}[t]{0.225\textwidth}
\centering
\resizebox{0.95\textwidth}{0.95\textwidth}{
\begin{tikzpicture}
\draw[-latex] (-1,-1,0) -- (-1,3.2,0);
\draw[-latex] (-1,-1,0) -- (5.2,-1,0);
\draw[-latex] (-1,-1,0) -- (-1,-1,-4.2);
\node at (-1.2,1.5,0) {\large{$\theta$}};
\node at (2,-1.2,0) {\large{$\mathit{t}$}};
\node[rotate=45] at (-.3,0,0) {{echoes}};

\node at (0.2,0,-.3) {$\bullet$};

\node at (0.4,0.8,0) {$\bullet$};
\node at (0.4,0.8,-.3) {$\bullet$};
\node at (0.4,0.8,.3) {$\bullet$};

\node at (0.8,2.4,0) {$\bullet$};

\node at (1.8,0.4,0) {$\bullet$};
\node at (1.8,0.4,-.3) {$\bullet$};

\node at (2,1.2,0) {\textcolor{red!60!black}{$\bullet$}};
\node at (2,1.2,.3) {$\bullet$};

\node at (2.2,2,0) {$\bullet$};

\node at (2.4,2.8,0) {$\bullet$};

\node at (3.6,0.8,0) {$\bullet$};

\node at (3.8,1.6,0) {$\bullet$};
\node at (3.8,1.6,-.3) {$\bullet$};

\node at (4,2.4,0) {$\bullet$};
\node at (4,2.4,-.3) {$\bullet$};

\draw[-latex,red!60!black] (2,1.2,0) -- (0.4,0.8,0);
\draw[-latex,red!60!black] (2,1.2,0) -- (0.4,0.8,.3);
\draw[-latex,red!60!black] (2,1.2,0) -- (0.4,0.8,-.3);
\draw[-latex,red!60!black] (2,1.2,0) -- (1.8,0.4,0);
\draw[-latex,red!60!black] (2,1.2,0) -- (1.8,0.4,-.3);
\draw[-latex,red!60!black] (2,1.2,0) -- (2.2,2,0);
\draw[-latex,red!60!black] (2,1.2,0) -- (3.6,0.8,0);
\draw[-latex,red!60!black] (2,1.2,0) -- (3.8,1.6,0);
\draw[-latex,red!60!black] (2,1.2,0) -- (3.8,1.6,-.3);
\end{tikzpicture}}
\caption{Echo sensor topology: each echo is connected to all echoes of the 6 neighboring pulses}
\label{fig:multi-echo}
\end{subfigure}
\caption{Definition of neighborhood in sensor space. For each figure, the points considered is colored in red, and connection is denoted by a red arrow.}
\end{figure}

\subsection{Reconstruction}
\label{sec:rec}

We improve the reconstruction presented in \citep{guinard2018sensor}, which was based on the following principles:
\begin{itemize}
\item $C_0$ regularity: we want to prevent forming edges between echoes when their euclidean distance is too high.
\item $C_1$ regularity: we want to favor edges when two collinear edges share an echo.
\end{itemize}

\begin{figure}[t]
\centering
\begin{tikzpicture}

\draw[-latex,black,thin,dashed] (0,2) -- (0,0);
\draw[-latex,black,thin,dashed] (2.5,2) -- (2.5,-.8);
\draw[-latex,black,thin,dashed] (5,2) -- (5,-1);
\draw[-latex,black,thin,dashed] (-2.5,2) -- (-2.5,-.3);

\node at (0,0) {$\bullet$};
\node at (2.5,0) {$\bullet$};
\node at (2.5,-.8) {$\bullet$};
\node[red!60!black] at (5,-.2) {$\bullet$};
\node[red!60!black] at (5,.7) {$\bullet$};
\node[red!60!black] at (5,-1) {$\bullet$};
\node[blue!60!black] at (-2.5,-.3) {$\bullet$};
\node[blue!60!black] at (-2.5,.6) {$\bullet$};


\node at (0,-.3) {\scriptsize{$E_p^{1}$}};
\node at (2.2,-.3) {\scriptsize{$E_{p+1}^{1}$}};
\node at (2.2,-1.1) {\scriptsize{$E_{p+1}^{2}$}};
\node at (-2.1,-.5) {\scriptsize{$E_{p-1}^{2}$}};
\node at (-2.1,.8) {\scriptsize{$E_{p-1}^{1}$}};
\node at (4.6,.9) {\scriptsize{$E_{p+2}^{1}$}};
\node at (4.6,-.5) {\scriptsize{$E_{p+2}^{2}$}};
\node at (4.6,-1.2) {\scriptsize{$E_{p+2}^{3}$}};

\draw[-latex,black] (0,0) -- (2.5,0);
\draw[-latex,black,dotted,thin] (0,0) -- (2.5,-.8);
\draw[-latex,red!60!black] (2.5,0) -- (5,-.2);
\draw[-latex,red!60!black,dotted,thin] (2.5,0) -- (5,-1);
\draw[-latex,red!60!black,dotted,thin] (2.5,0) -- (5,.7);
\draw[-latex,blue!60!black] (-2.5,-.3) -- (0,0);
\draw[-latex,blue!60!black,dotted,thin] (-2.5,.6) -- (0,0);


\node at (1.25,.2) {\scriptsize{$\vec{e_p}(e_1,e_2)$}};
\node[rotate=5] at (-1.3,.05) {\scriptsize{$\vec{e_{p-1}}(e,e_1)$}};
\node[rotate=-5] at (3.8,.1) {\scriptsize{$\vec{e_{p+1}}(e_2,e)$}};
\node at (.3,1) {$\vec{l_p}$};

\end{tikzpicture}
\caption{Illustration of the computed regularities $C_0$ and $C_1$. The black dots represent the echoes associated to the considered pulses. The blue and red ones correspond respectively to the precedent and following adjacent echoes. The solid arrows show the adjacent echoes used for the $C_0$ and $C_1$ computation. The black one correspond to the most orthogonal liaison to the sensor beams. The blue and red ones are selected because the angle between these vectors and the black one are the closest possible to $\pi$. The black dashed lines represent the laser beams.}
\label{fig:regularities}
\end{figure}
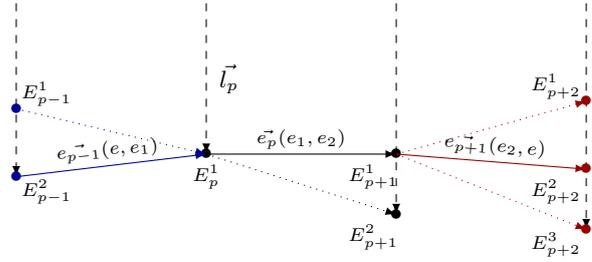

In order to be independent from the sampling, and to exploit the hexagonal structure of the sensor topology, both regularities were expressed in an angular manner and computed on every direction of the structure independently. For the reminder of this article, and because a single pulse can have multiple echoes, we will express, for a pulse ${p}$, its echoes as $E_p^{e}$ where ${e} \in 1\ldots N_p$, with ${N_p}$ the number of echoes of ${p}$. We expressed the regularities as follow:

\begin{itemize} 
\item $C_0$ regularity, for an edge  $(E_{p}^{e_1}, E_{p+1}^{e_2})$ between two echoes of  two neighboring pulses: $$C_0(\mathit{p}, e_1, e_2)= 1 - \vec{e_p}(e_1, e_2)\cdot\vec{l_p} \quad,$$ where $ \vec{e_p}(e_1, e_2) = \frac{\overrightarrow{E_{p}^{e_1}E_{p+1}^{e_2}}}{||\overrightarrow{E_{p}^{e_1}E_{p+1}^{e_2}}||}$ and $\vec{l_p}$ is the direction of the laser beam of pulse $p$ (cf Figure \ref{fig:regularities}). $C_0$ is close to 0 for surfaces orthogonal to the LiDAR ray and close to 1 for grazing surfaces, almost parallel to the ray.
\item $C_1$ regularity, for an edge  $(E_{p}^{e_1}, E_{p+1}^{e_2})$ between two echoes of  two neighboring pulses:
\begin{equation}
\nonumber
\begin{split}
C_1(p, e_1, e_2) \!\!=\!\,& min_{e=1}^{N_{p-1}} |1\!\!-\!\vec{e}_{p-1}(e, e_1)\!\cdot\! \vec{e}_p(e_1, e_2)|\cdot \\ 
& min_{e=1}^{N_{p+2}} |1\!\!-\!\vec{e}_p(e_1, e_2)\!\cdot\! \vec{e}_{p+1}(e_2, e)|. 
\end{split}
\end{equation}
where the minima are given a value of 1 if the pulse is empty. $C_1$ is close to 0 is the edge is aligned with at least one of its neighboring edges, and close to 1 if it is orthogonal to all neighboring edges.
\end{itemize}

From the $C_0$ regularity,
we derive the following weighted $C_0^{w}$ regularity:
$$
C_0^w(p,e_1,e_2) = C_0(p,e_1,e_2) + \kappa \cdot \frac{{l_p}}{{l_{max}}},
$$
where $l_p$ is the distance from the sensor to the point, ${l_{max}}$ is the maximum distance between a position of a laser and one of its recorded echoes and $\kappa$ is the parameter conditioning the influence of the weighting term. The $C_0^w$ is no longer independent from the sampling and varies between $0$ and $1 + \kappa$. 

The performances of this parameter can also be improved by adding a threshold on edge length. This way, we prevent the apparition of long edges between objects far away from one another (e.g. with a depth difference of several meters). We propose this approach because we find very unlikely the case where an edge of a few meters long will be part of the reality. It would mean that there was an object in the scene that was big enough to have a few meters side, and that this object is very close to be parallel to the laser beam. Intuitively, this description could correspond to a building, but actually, there are very little points behind the main facade of a building and they are too sparse to let us find a possible shape.

\section{Results}

We implemented the pipeline presented before, first without the weighting. Then, we added the weighting formulation. We compared the results of both methods.

For all the following tests, we used data from the Stereopolis vehicle \citep{paparoditis2012stereopolis}. The scenes have been acquired in an urban environment (Paris). All the simplicial complexes presented in this part will be represented as follow:
\begin{itemize}
\item triangles in red,
\item edges that are not part of any triangle in green,
\item points that don't belong to any triangle or edge in black.
\end{itemize}
Note that following its mathematical definition, the endpoints of an edge of a simplicial complex also belong to the complex, and similarly for the edges of a triangle, but we do not display them for clarity.

We first study the influence of $\kappa$ with and without a threshold on edge length. Then we compare our method to the method presented in \citep{guinard2018sensor}.


\subsection{Parametrization of $\kappa$}

In this section, we investigate the influence of the weighting parameter $\kappa$ on the reconstruction. We conducted two different sets of experiments: the first one shows the influence of $\kappa$ if we do not threshold edge length. The second one corresponds to the case where we limited edge maximum length, thus enabling the reconstruction of much more triangles and edges.

\begin{figure*}[t]
\centering
\begin{subfigure}[t]{0.23\textwidth}
\centering
\includegraphics[width=1\textwidth]{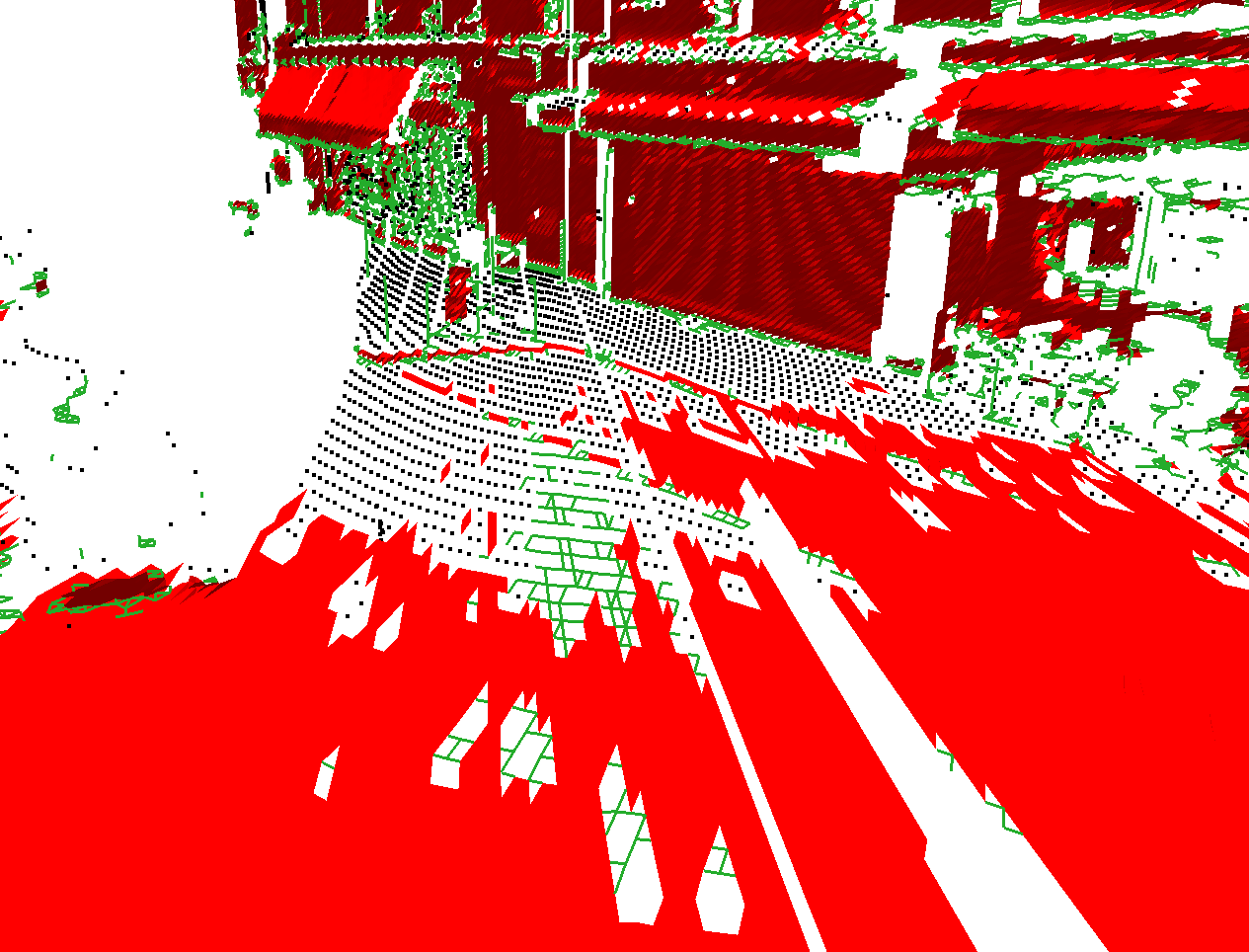}
\caption{$\kappa$= 0}
\end{subfigure}
\hspace{.2cm}
\begin{subfigure}[t]{0.23\textwidth}
\centering
\includegraphics[width=1\textwidth]{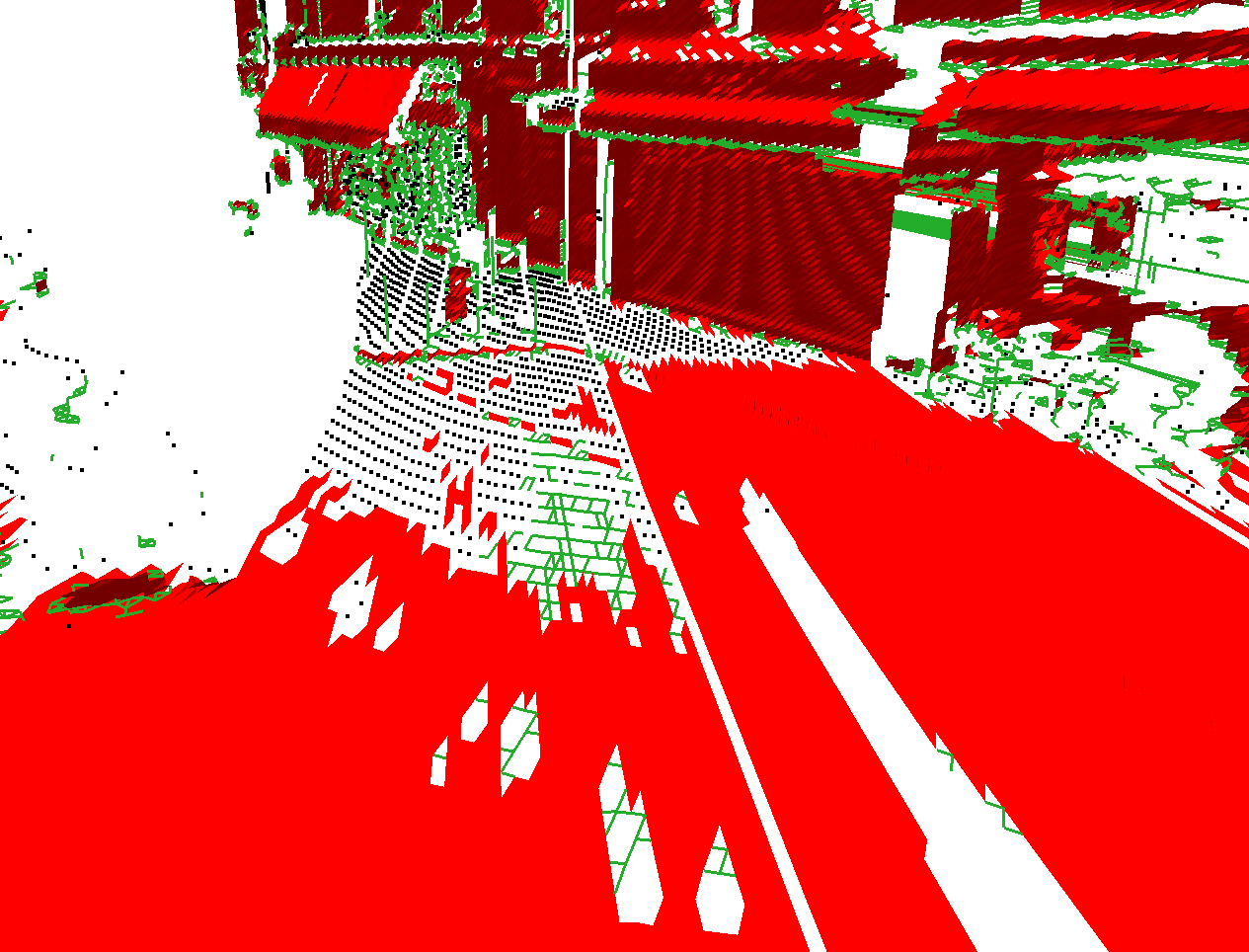}
\caption{$\kappa$= 0.1}
\end{subfigure}
\hspace{.2cm}
\begin{subfigure}[t]{0.23\textwidth}
\centering
\includegraphics[width=1\textwidth]{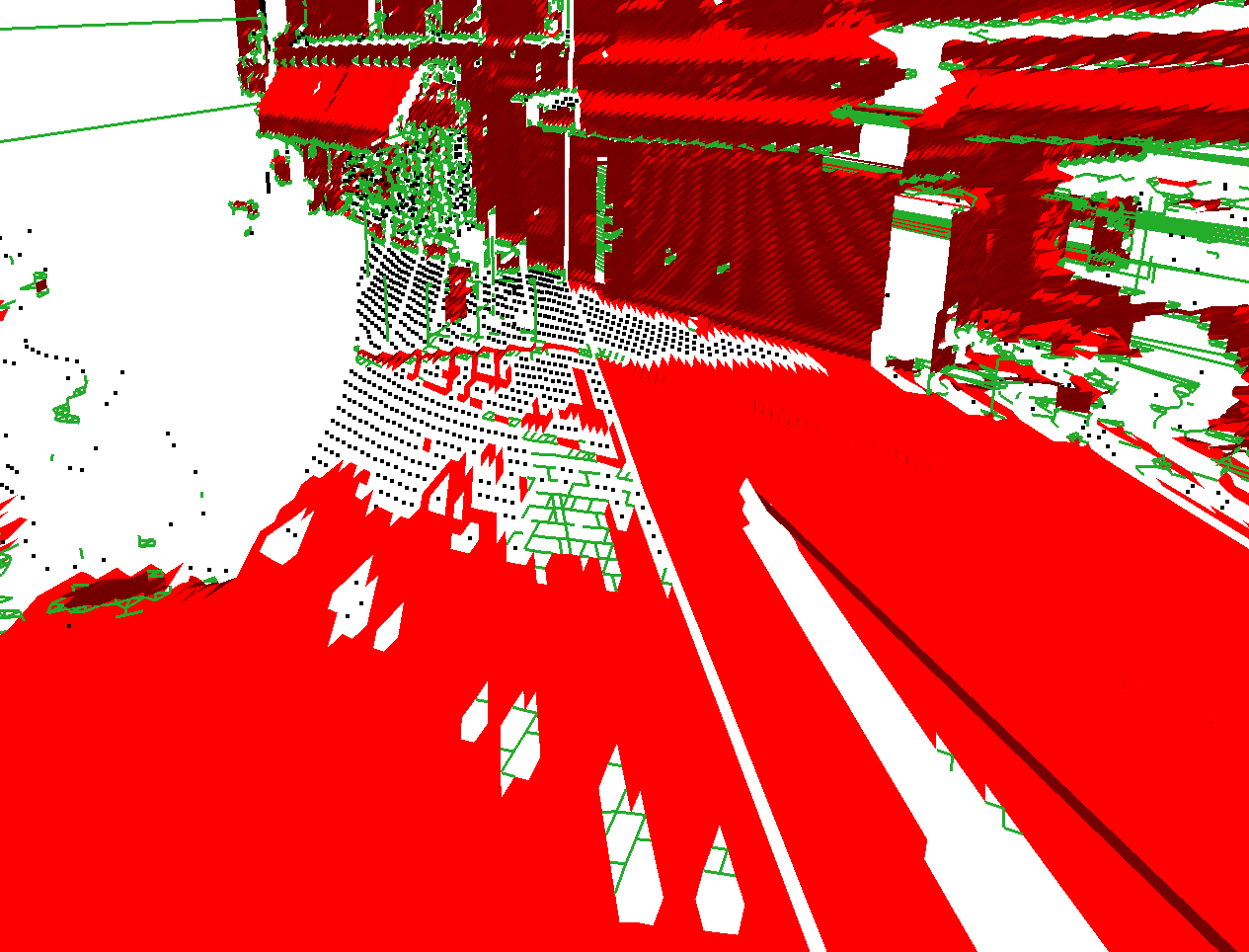}
\caption{$\kappa$= 0.2}
\end{subfigure}
\hspace{.2cm}
\begin{subfigure}[t]{0.23\textwidth}
\centering
\includegraphics[width=1\textwidth]{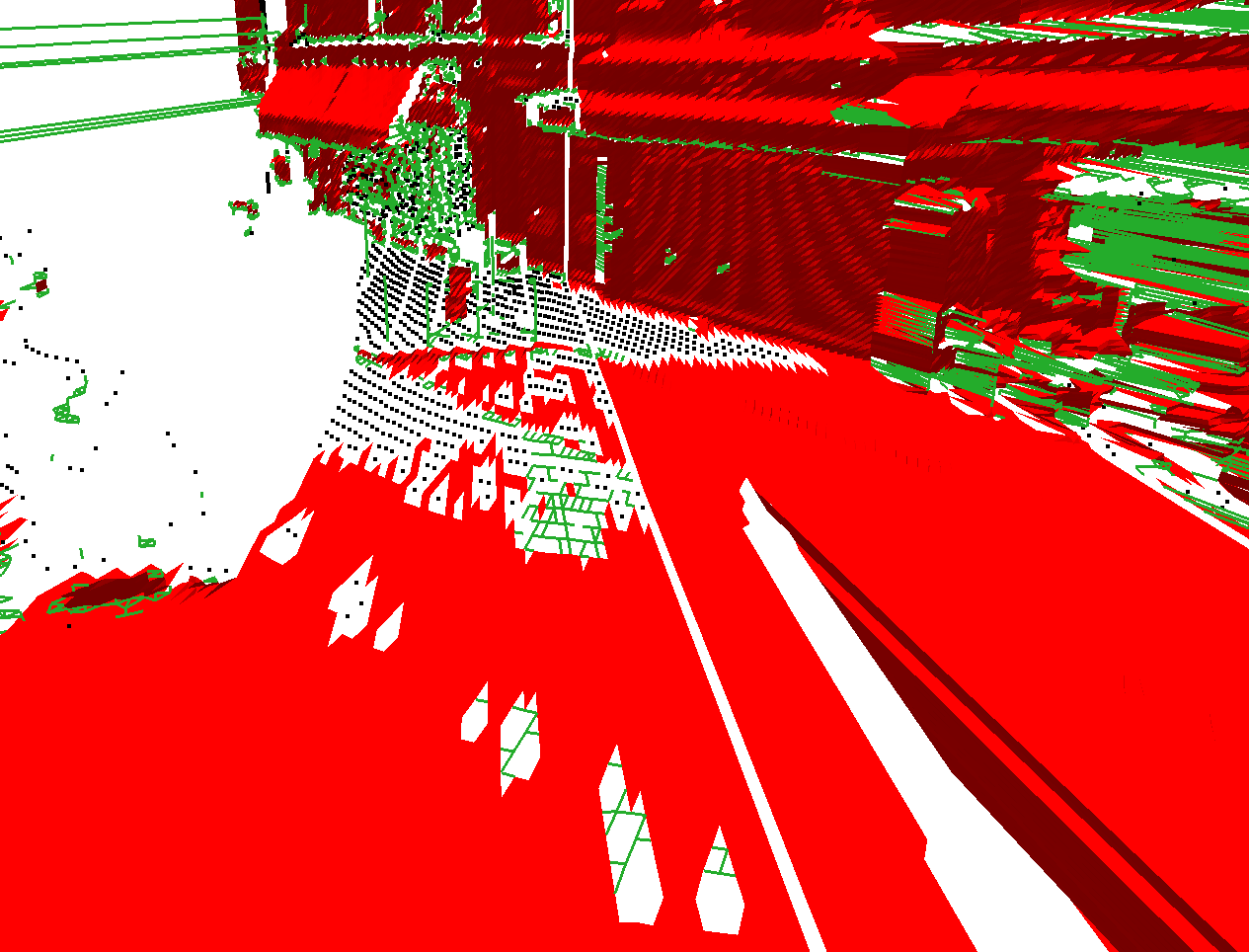}
\caption{$\kappa$= 0.3}
\end{subfigure}
\caption{Influence of $\kappa$. The scene represents a road far from the laser, with a facade in the background. The main differences between the images can be seen on the upper part of the road, which is slowly filled as $\kappa$ increases; and on the edges between facades and poles or pedestrians.}
\label{fig:kappa}
\end{figure*}

For this set of experiments, the values of $\alpha_m$, $\lambda$, $\omega$ and $\epsilon$ where respectively fixed to: $5 \cdot 10^{-2}$, $10^{-3}$, $10^{-3}$ and $5\cdot 10^{-3}$. We expect the $C_0^w$ regularity to improve the reconstructions using only the $C_0$ regularity, by allowing a few edges in the simplicial complex that where discarded otherwise. These edges should also encourage the creation of triangles on places that contained holes, and in the farthest places of the scene. The results are shown on figure \ref{fig:kappa}. 
The top left figure is the output produced using \citep{guinard2018sensor}. We clearly see that the weighting of the $C_0$ regularity helps our algorithm retrieve edges and triangles that were lost before. Low values of $\kappa$ fulfill some holes of the road, whereas high values of $\kappa$ create edges between objects far away one from another.
The number of points, edges and triangles for each simplicial complex is shown in table \ref{tab:repartition1}. There is a significant decrease of the number of points and egdes when $\kappa$ rises, whereas the number of triangles increases a bit.

\begin{center}
\begin{table}[h]
\centering
\begin{tabular}{c|ccc}
 & Triangles & Edges & Points \\
 \hline
$\kappa = 0$ & 737596 & 364730 & 12090 \\
$\kappa = 0.1$ & 744702 & 353968 & 10944 \\
$\kappa = 0.2$ & 752726 & 333926 & 10331 \\
$\kappa = 0.3$ & 758714 & 348742 & 9076 \\
\end{tabular}
\caption{Number of triangles, edges and points per simplicial complex in figure \ref{fig:kappa}.}
\label{tab:repartition1}
\end{table}
\end{center}

\begin{figure*}[t]
\centering
\begin{subfigure}[t]{0.23\textwidth}
\centering
\includegraphics[width=1\textwidth]{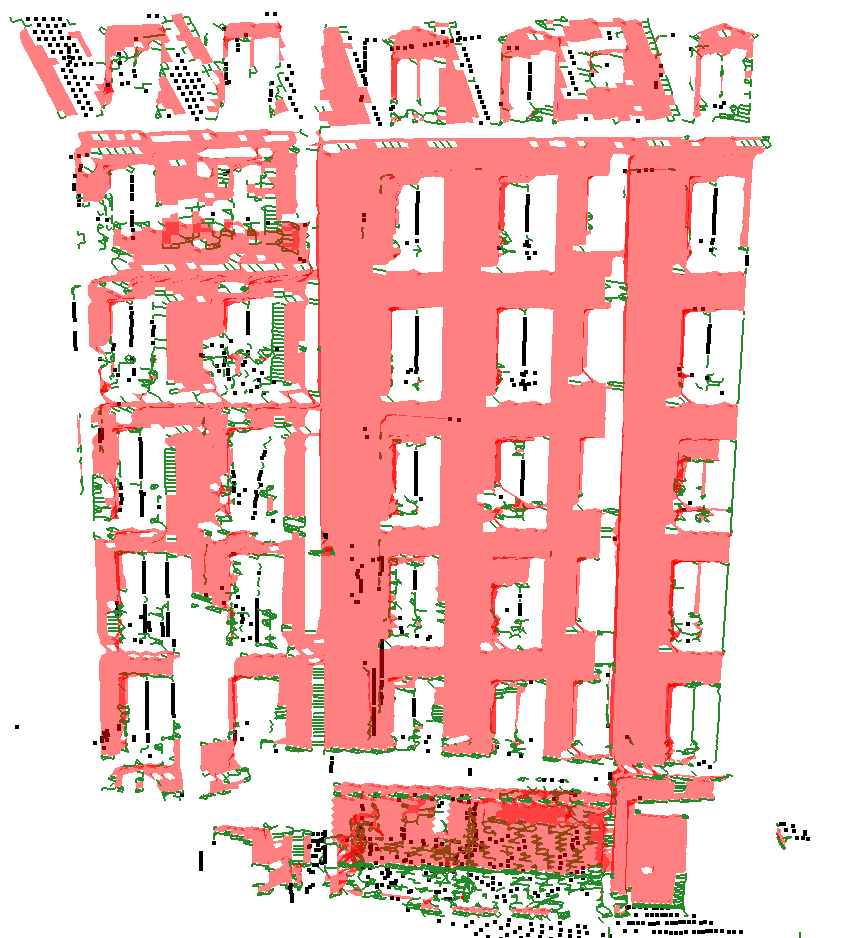}
\caption{$\kappa$= 0 and no edge length filtering}
\end{subfigure}
\hspace{.2cm}
\begin{subfigure}[t]{0.23\textwidth}
\centering
\includegraphics[width=1\textwidth]{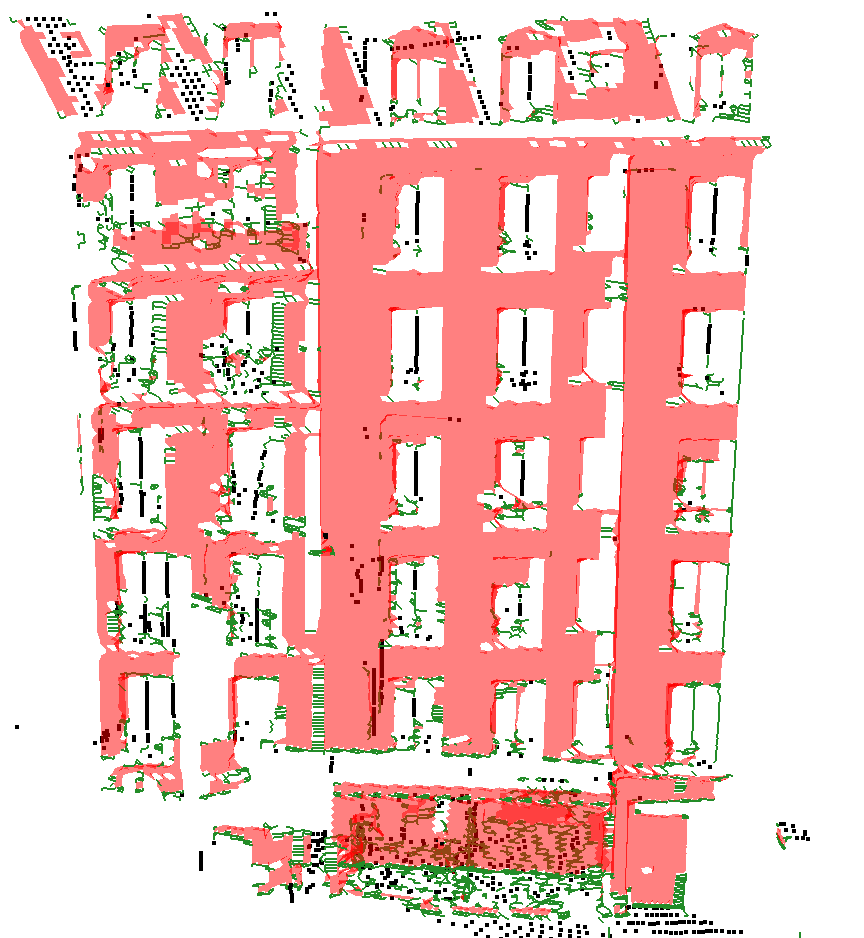}
\caption{$\kappa$= 0.03 and no edge length filtering}
\end{subfigure}
\hspace{.2cm}
\begin{subfigure}[t]{0.23\textwidth}
\centering
\includegraphics[width=1\textwidth]{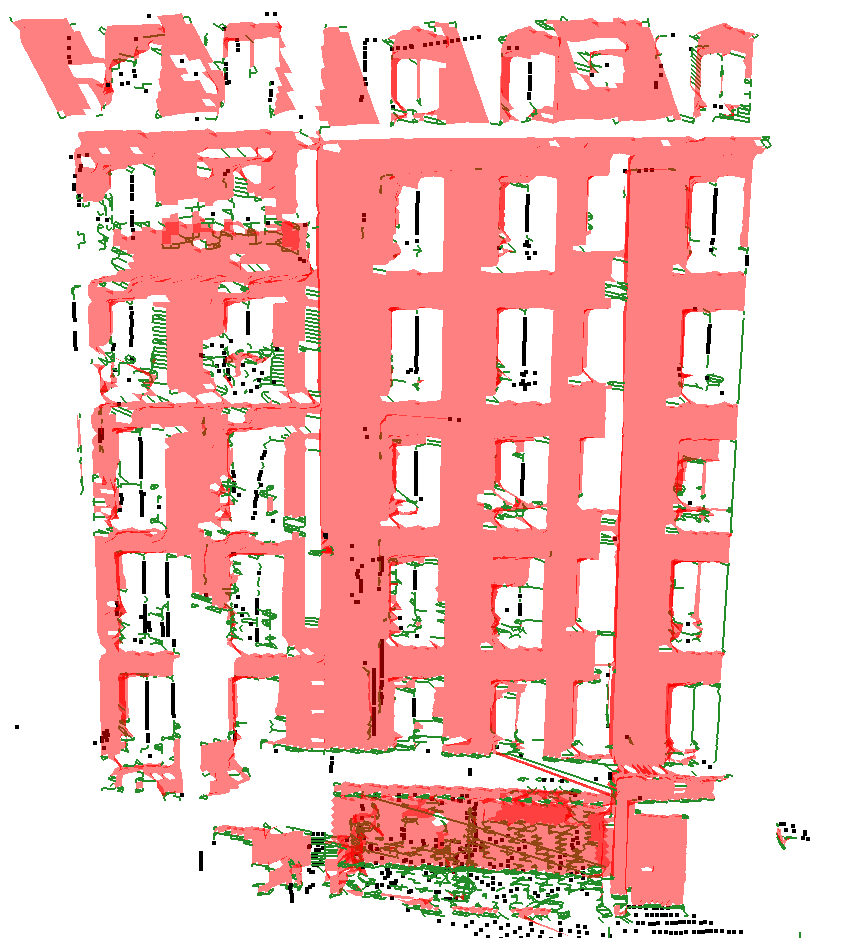}
\caption{$\kappa$= 0.2 and edge length filtering of 10 meters}
\label{fig:kappa_tresh_02}
\end{subfigure}
\hspace{.2cm}
\begin{subfigure}[t]{0.23\textwidth}
\centering
\includegraphics[width=1\textwidth]{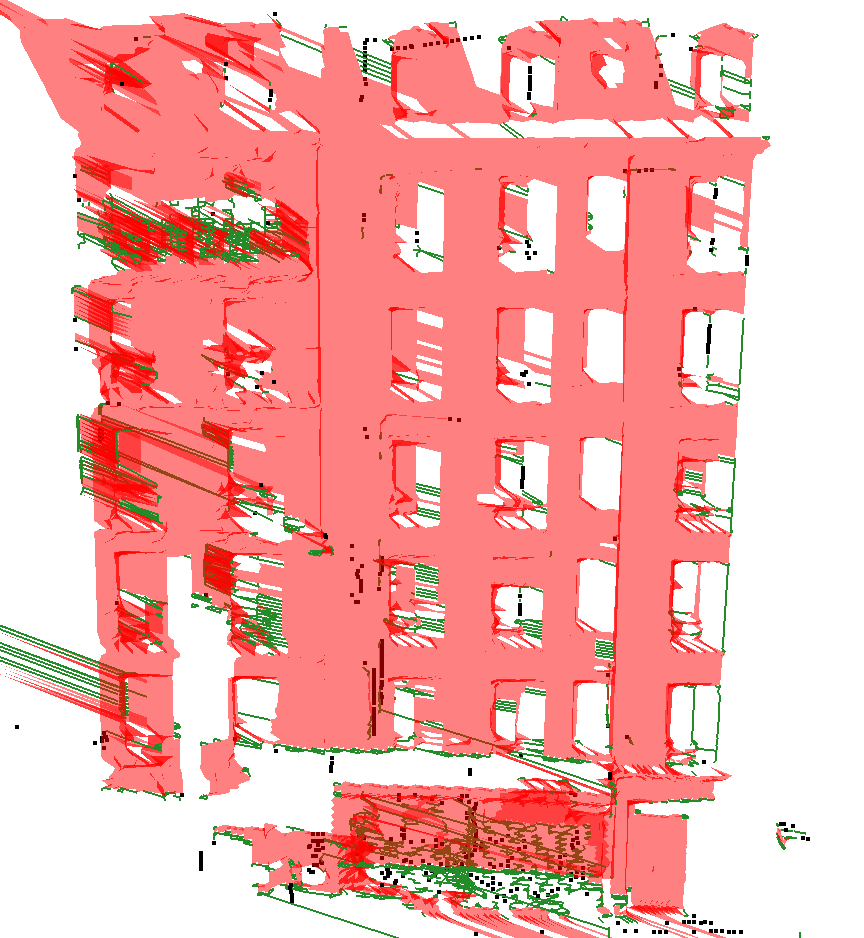}
\caption{$\kappa$= 0.5 and edge length filtering of 10 meters}
\end{subfigure}
\caption{Influence of $\kappa$. The scene represents a facade far from the laser.}
\label{fig:kappa_tresh}
\end{figure*}

The results of the second set of experiments are presented on figure \ref{fig:kappa_tresh}. Here we wanted to study the influence of $\kappa$ when we prevent the reconstruction of drawn-out edges. We fixed a maximum length for all edges to 10 meters. The figure on the left corresponds to a reconstruction without any weighting. Next, the weighting without any thresholding is done for $\kappa = 0.03$. Last we show two examples of thresholding edges longer than 10 meters with a $\kappa$ of respectively $0.2$ and $0.5$. Using this thresholding allows to use a higher value of $\kappa$, thus creating a cleaner, less holed reconstruction, especially visible in figure \ref{fig:kappa_tresh_02}, even if a too high value of $\kappa$ keeps creating edges between objects not connected in the scene. 

\subsection{Comparison of both methods}

In this part, we compare our method to the method presented in \citep{guinard2018sensor}. For these methods,  $\alpha_m$, $\lambda$, $\omega$, $\epsilon$ and $\kappa$ are respectively fixed to $0.05$, $10^{-4}$, $0.1$, $5\cdot10^{-3}$ and $0.4$.

\begin{figure*}[t]
\centering
\begin{subfigure}[t]{0.49\textwidth}
\centering 
\includegraphics[width=.97\textwidth]{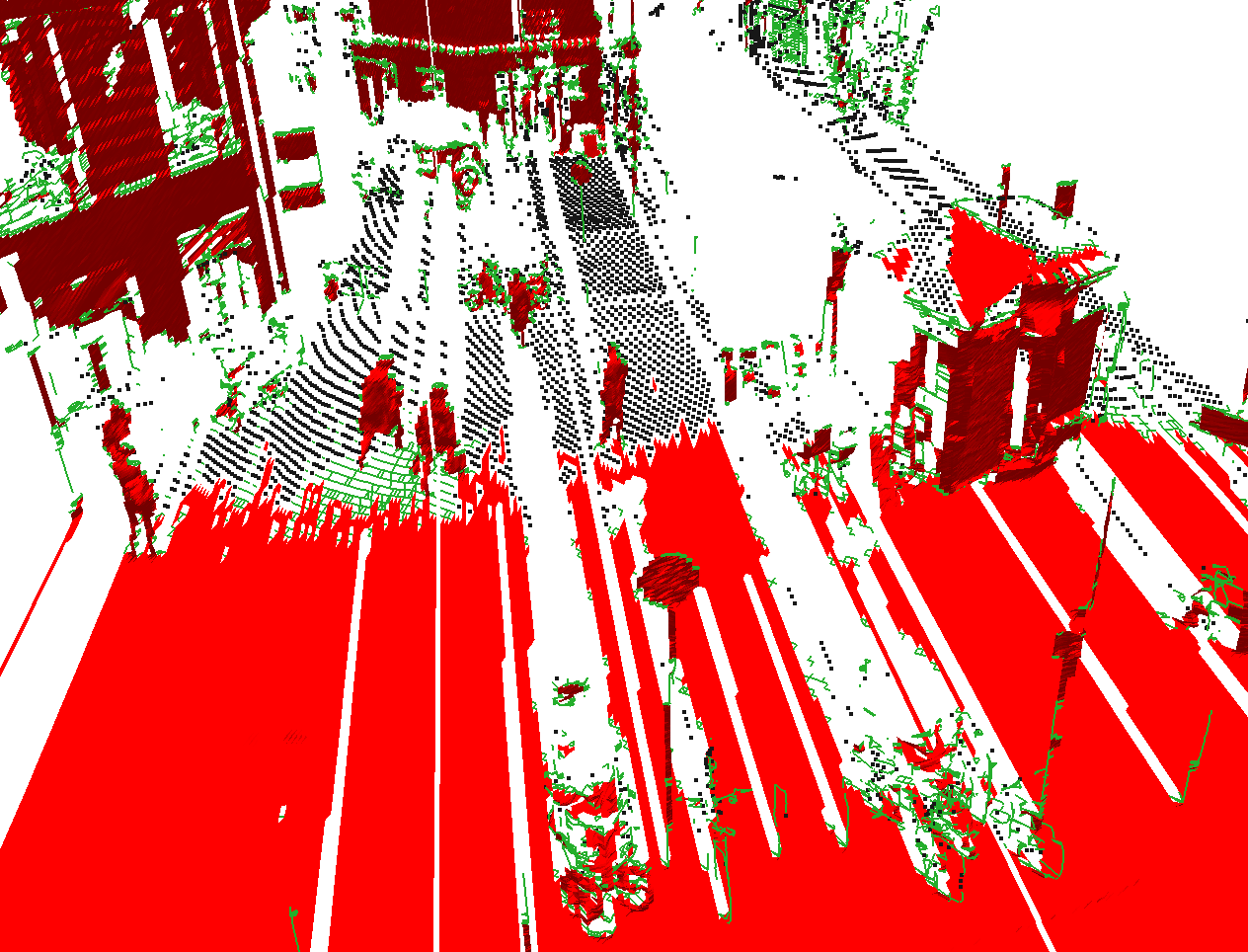}
\caption{Unweighted}
\end{subfigure}
\begin{subfigure}[t]{0.49\textwidth}
\centering
\includegraphics[width=.97\textwidth]{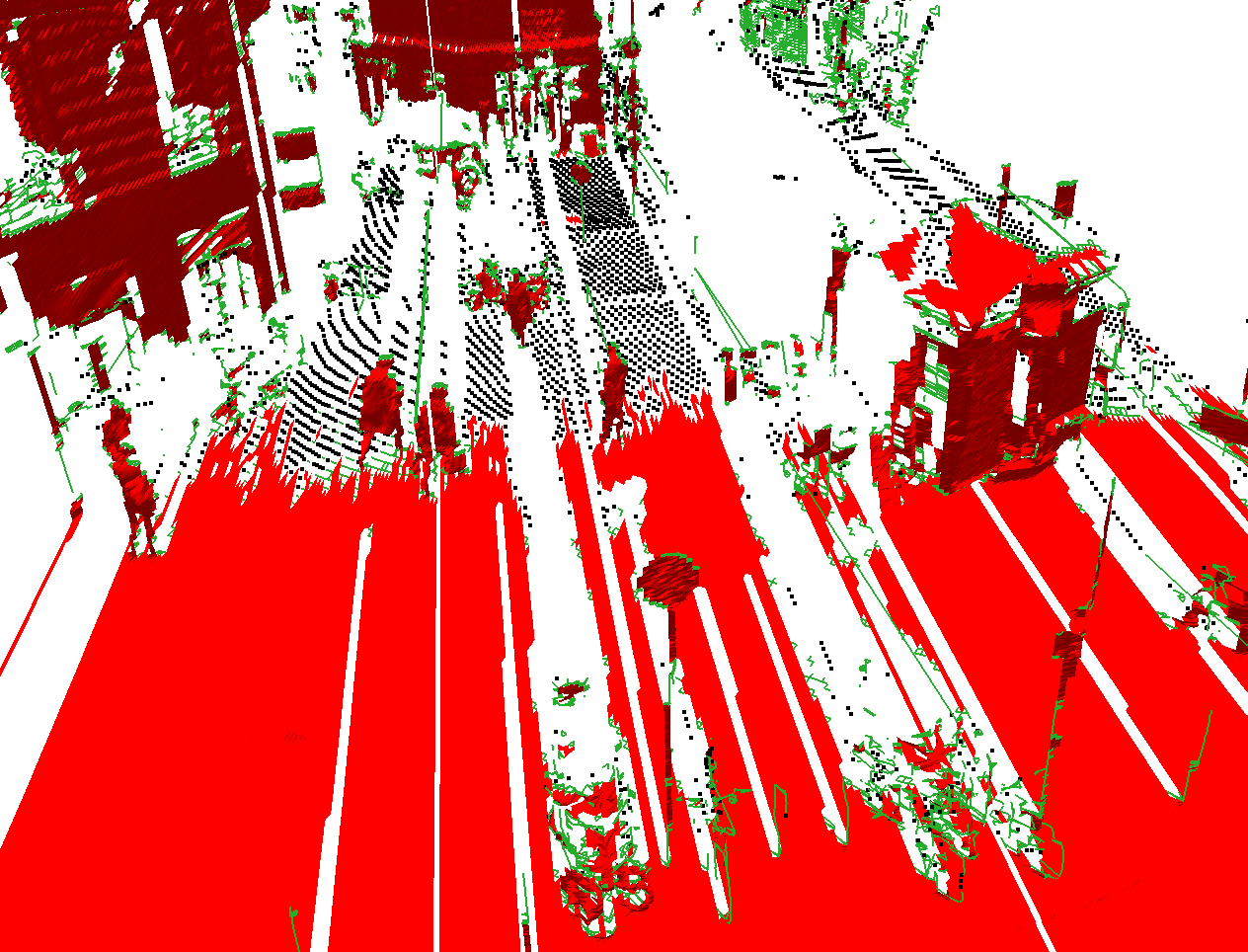}
\caption{Weighted}
\label{fig:urban_weighted}
\end{subfigure}
\caption{Results on a complete urban scene, with road, facades, poles and pedestrians.}
\label{fig:urban}
\end{figure*}

\begin{figure*}[!t]
\centering
\begin{subfigure}[t]{1\textwidth}
\centering
\setlength{\tabcolsep}{3pt} 
\renewcommand{\arraystretch}{.7} 
\begin{tabular}{cccc} 
\includegraphics[scale=.12]{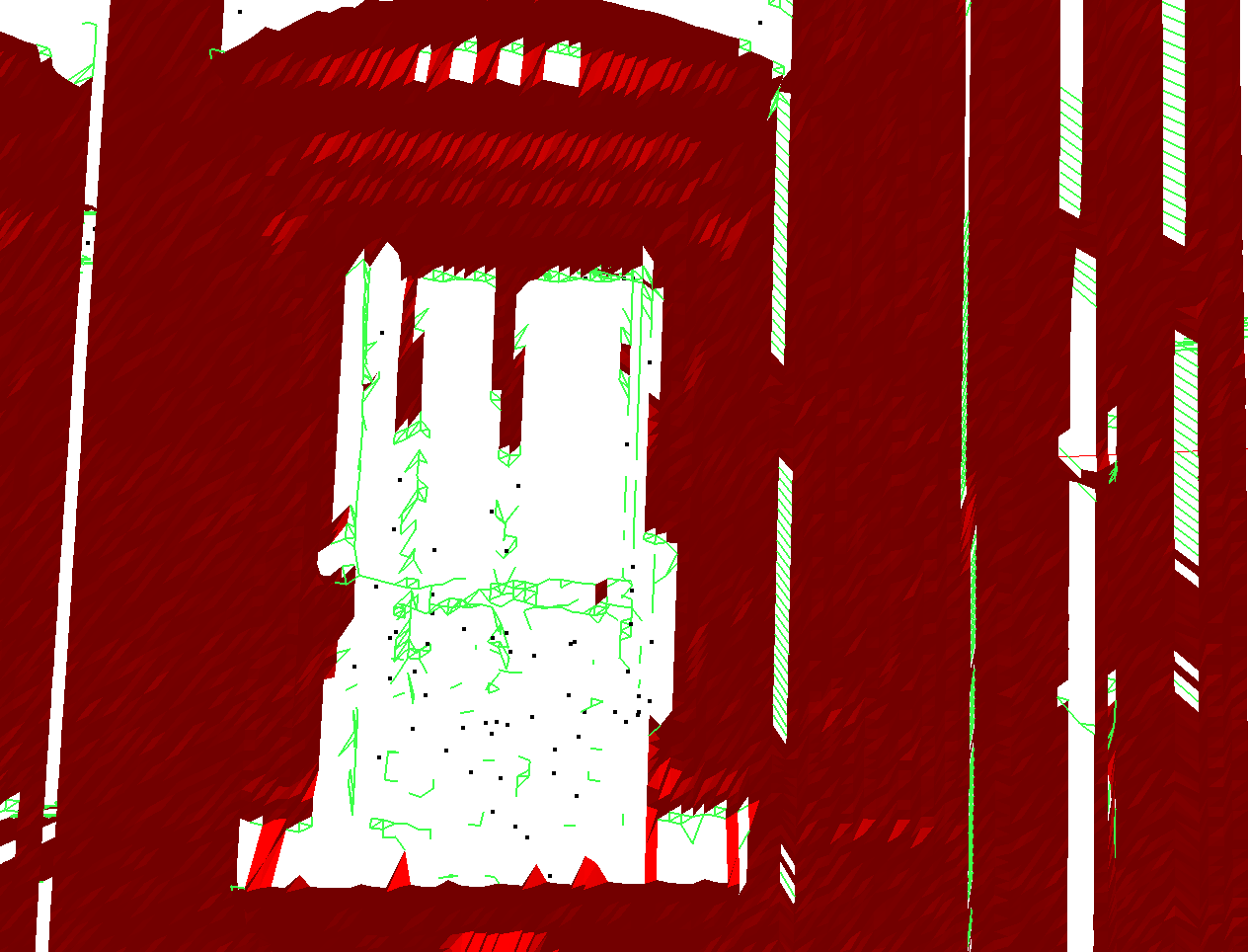} &
\includegraphics[scale=.12]{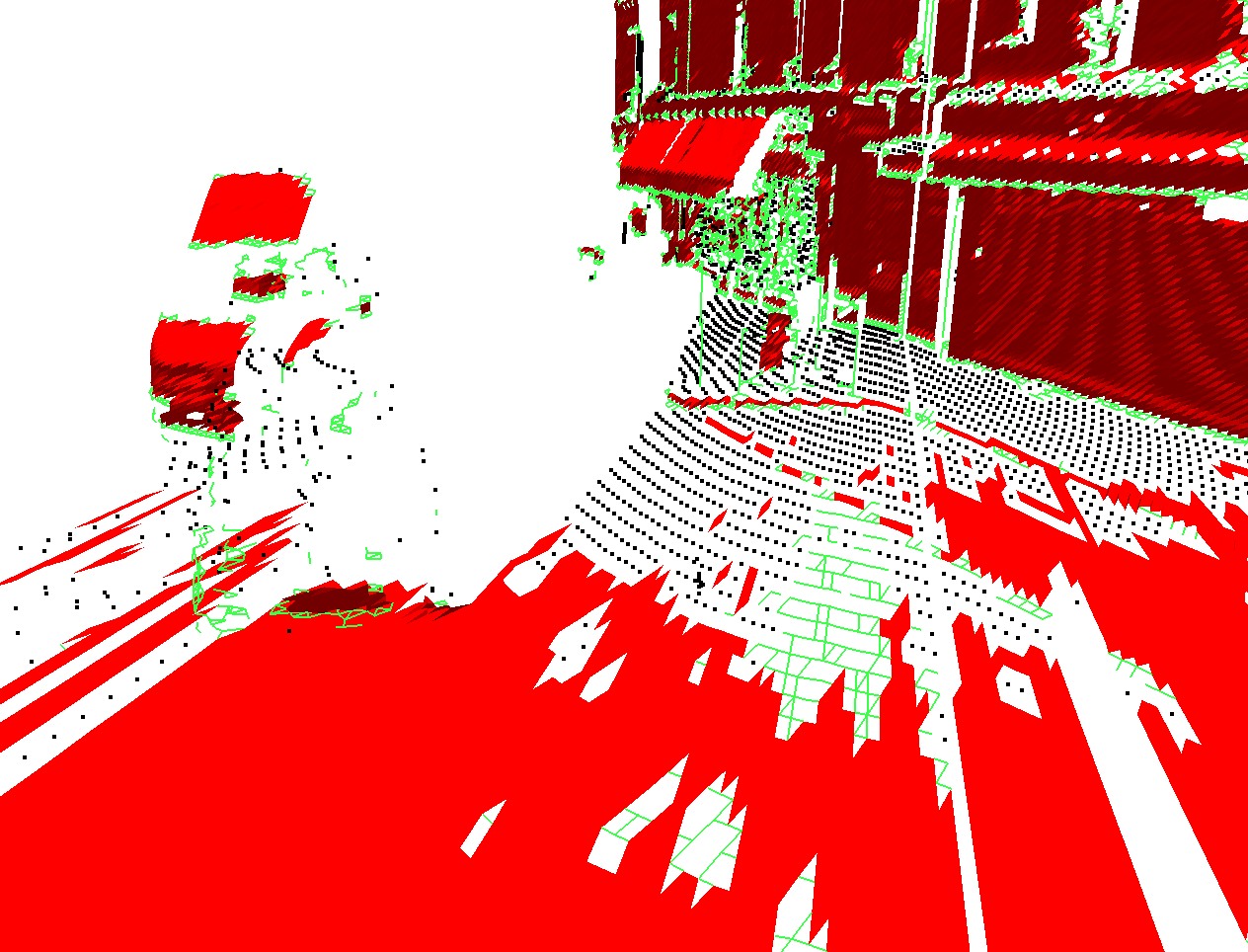} &
\includegraphics[scale=.12]{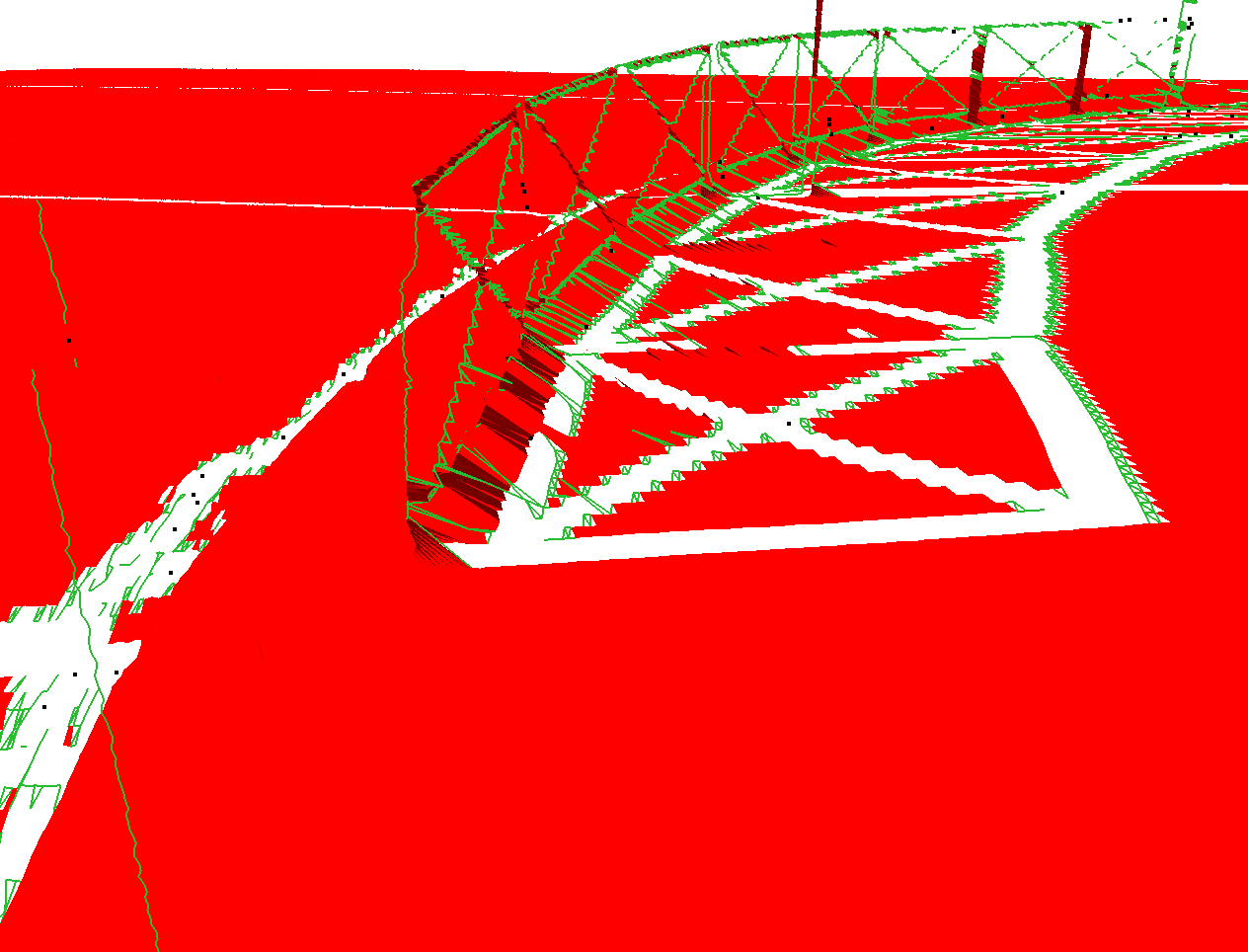} 
\end{tabular}
\caption{Unweighted reconstruction \citep{guinard2018sensor}}
\end{subfigure}

\begin{subfigure}[t]{1\textwidth}\vspace{.3cm}
\centering
\setlength{\tabcolsep}{3pt} 
\renewcommand{\arraystretch}{.7} 
\begin{tabular}{cccc}
\includegraphics[scale=.12]{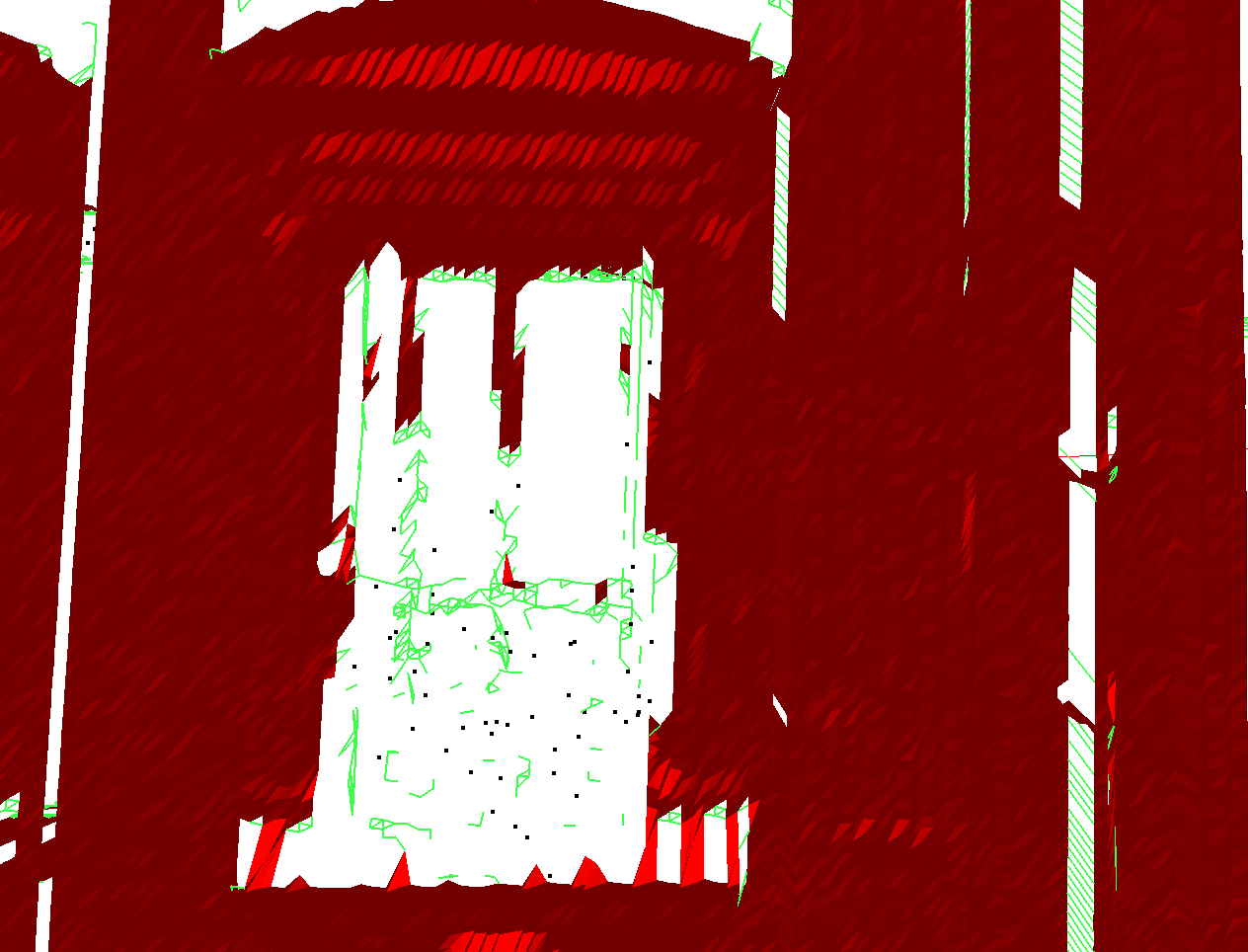} &
\includegraphics[scale=.12]{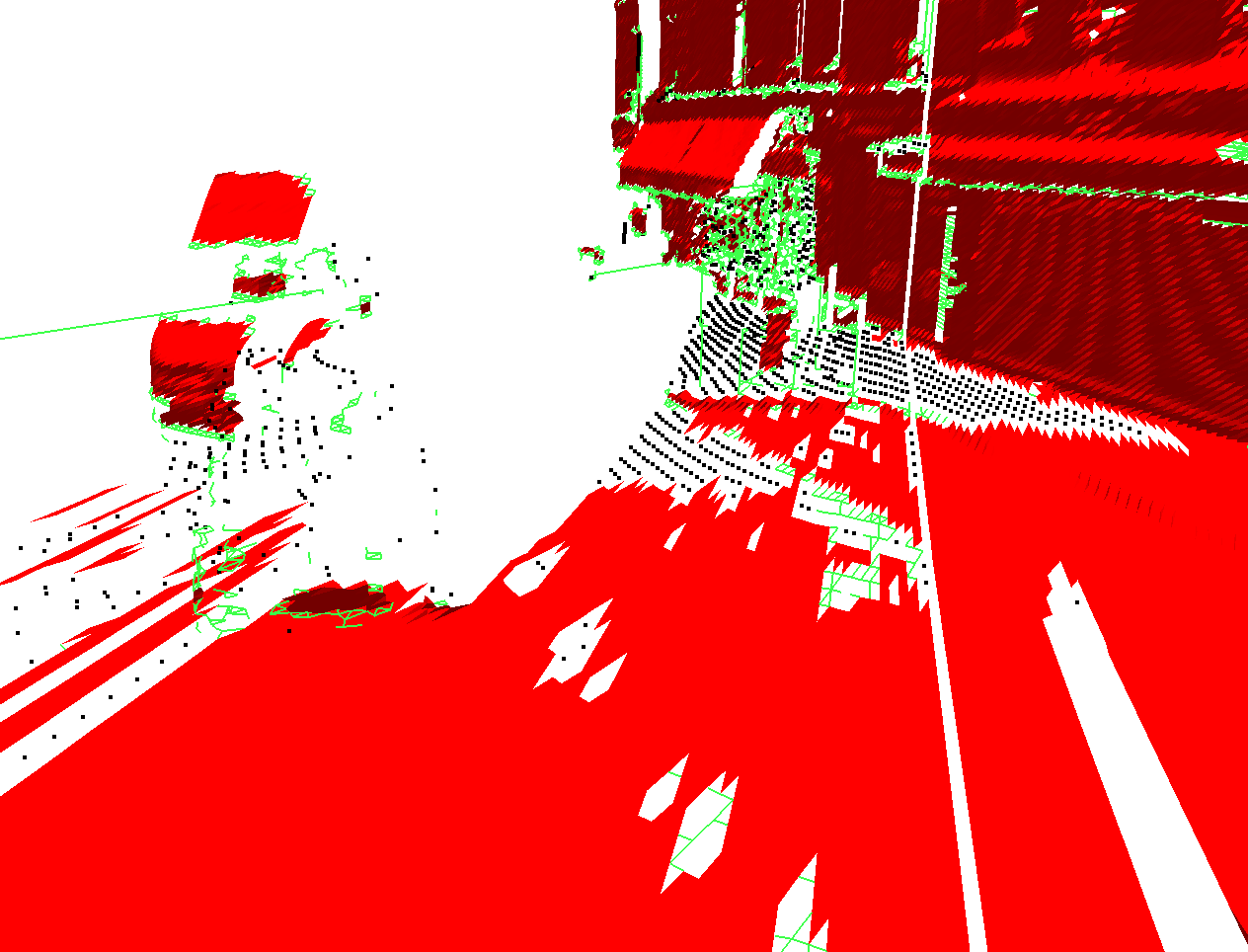} &
\includegraphics[scale=.12]{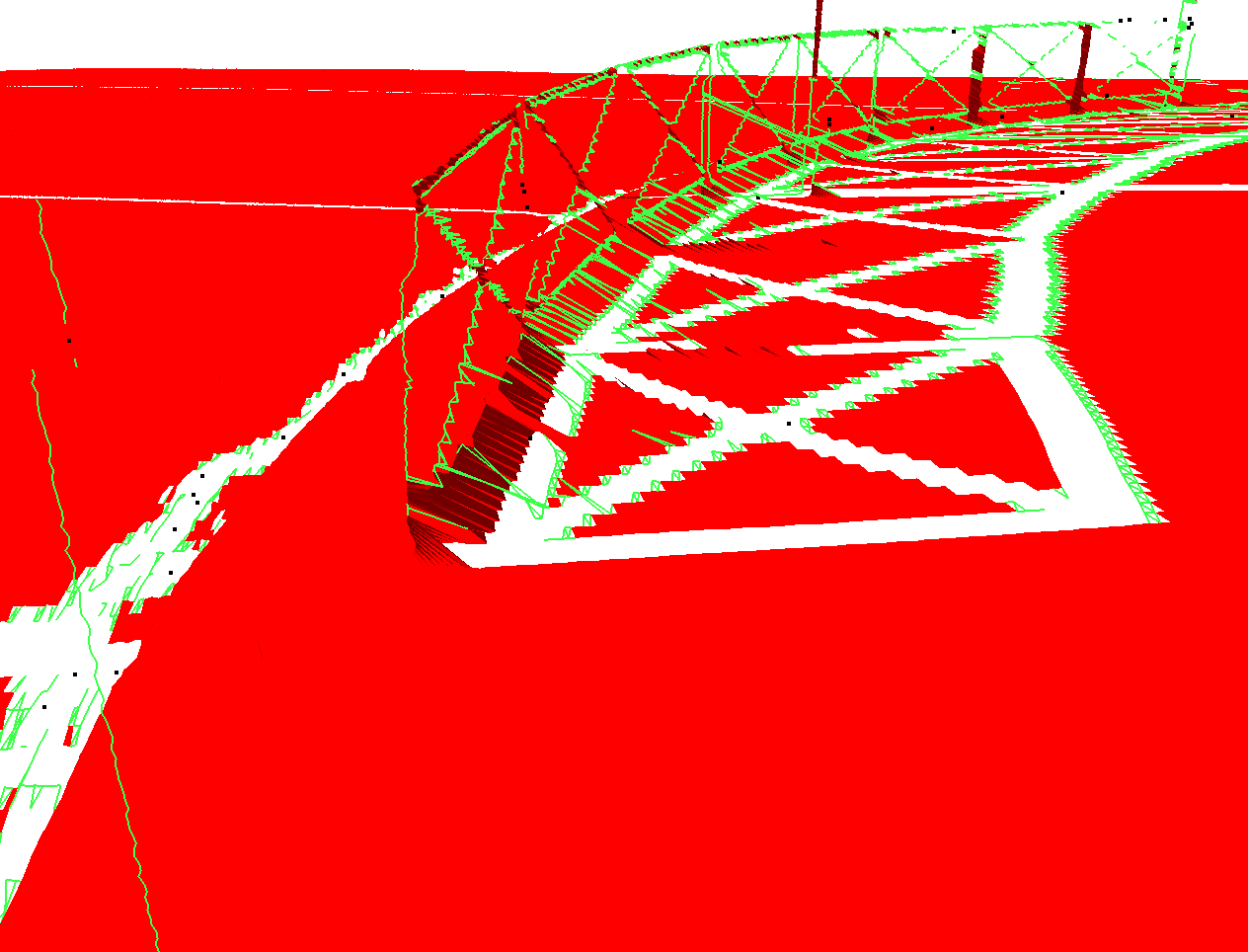} 
\end{tabular}
\caption{Weighted reconstruction}
\end{subfigure}
\caption{Comparison of the two methods: the unweighted and the weighted methods. From left to right, the scenes represent: a window, a road in the grazing surface case and fences.}
\label{fig:comparison}
\end{figure*}

Figure \ref{fig:urban} presents a reconstruction in a complex urban scene with facades, roads, trees $\ldots$. The figure on the left shows the results of \citep{guinard2018sensor} and the right one ours. There is nearly no difference on small objects like poles or pedestrians. 
Moreover, we can see that our method is able to retrieve more surfaces in the limits of laser's scope. This is showned by filled facades in the top of the image, and also by a cleaner road, even if the whole reconstruction of the road would recquire a higher value of $\kappa$ that would just spoil the remaining parts of the reconstruction.
A zoom on specific areas of the scene (windows, fences $\ldots$) is visible on figure \ref{fig:comparison}. We note that on parts of the cloud were the reconstruction without weighting performed well, the add of weight does not harm the results. Most differences between both reconstructions happen on grazing surfaces, where our method is more efficient.

The white lines visible in some figures corresponds to occlusions or limits between sections of one second of acquisition from the MLS. 
The number of triangles, edges and points are stored in table \ref{tab:repartition2}. Again, the number of points and edges dicreases and the number of triangles rises a bit. This is mainly due to the fact that by authorizing more edges in the first step of the reconstruction, our method can produce more triangles later, thus dicreasing the number of remaining edges and points.

\begin{center}
\begin{table}[h]
\centering
\begin{tabular}{c|ccc}
 & Triangles & Edges & Points \\
 \hline
Unweighted & 1143482 & 755582 & 13757 \\
Weighted & 1153932 & 713064 & 11942 \\
\end{tabular}
\caption{Number of triangles, edges and points per simplicial complex in figure \ref{fig:urban}.}
\label{tab:repartition2}
\end{table}
\end{center}

\section{Conclusions and perspectives}

This article presented a method for the reconstruction of simplicial complexes of point clouds from MLS, based on the inherent structure of the MLS. We propose a filtering of edges possibly linking adjacent echoes by searching for collinear edges in the cloud, or edges perpendicular to the laser beams. We added a weighting parameter to this formulation in order to produce a simplicial complex less holed, especially on areas far from the sensor, where points are naturally further one from another.

The main drawback of this method is its propensity to create very long edges between objects not connected in real life. Even it is partially corrected by thresholding edges length, allowing to compute a reconstruction with higher values of $\kappa$ it does not fully retrieve edges on objects far from the sensor.

Further developments may also consider a hole filling process, to get rid of the absence of a few missing simplexes in a large structure (road, building) as in \citep{harary2014context}. Moreover, setting up a generalization method, like in \citet{popovic1997progressive}, would be interesting to simplify the resulting simplicial complexes on large and regular structures in order to reduce the memory weight of the simplicial complexes while maintaining a high accuracy. Last, our reconstruction may be useful to help a segmentation algorithm in order to obtain a semantic segmentation of the scene \citep{rusu2009close,jeong2018multimodal}.

\section*{Acknowledgments}

The authors would like to acknowledge the DGA for their financial support of this work.

\renewcommand{\bibname}{Bibliography}
\renewcommand{\refname}{Bibliography}
\bibliographystyle{rfpt}
\bibliography{CFPT2018}

\begin{thebibliography}{21}
\expandafter\ifx\csname natexlab\endcsname\relax\def\natexlab#1{#1}\fi
\expandafter\ifx\csname url\endcsname\relax
  \def\url#1{\texttt{#1}}\fi
\expandafter\ifx\csname urlprefix\endcsname\relax\def\urlprefix{URL }\fi

\bibitem[{Alliez et~al.(2007)Alliez, Cohen-Steiner, Tong, and
  Desbrun}]{alliez2007voronoi}
Alliez, P., Cohen-Steiner, D., Tong, Y., Desbrun, M., 2007. Voronoi-based
  variational reconstruction of unoriented point sets. In: Symposium on
  Geometry processing. Vol.~7. pp. 39--48.

\bibitem[{Becker and Haala(2009)}]{becker2009grammar}
Becker, S., Haala, N., 2009. Grammar supported facade reconstruction from
  mobile lidar mapping. In: ISPRS Workshop, CMRT09-City Models, Roads and
  Traffic. Vol.~38. p.~13.

\bibitem[{Berger et~al.(2014)Berger, Tagliasacchi, Seversky, Alliez, Levine,
  Sharf, and Silva}]{berger2014state}
Berger, M., Tagliasacchi, A., Seversky, L., Alliez, P., Levine, J., Sharf, A.,
  Silva, C., 2014. State of the art in surface reconstruction from point
  clouds. In: EUROGRAPHICS star reports. Vol.~1. pp. 161--185.

\bibitem[{Bernardini and Bajaj(1997)}]{bernardini1997sampling}
Bernardini, F., Bajaj, C.~L., 1997. Sampling and reconstructing manifolds using
  alpha-shapes. In: In Proc. 9th Canad. Conf. Comput. Geom. Citeseer.

\bibitem[{De~Goes et~al.(2011)De~Goes, Cohen-Steiner, Alliez, and
  Desbrun}]{de2011optimal}
De~Goes, F., Cohen-Steiner, D., Alliez, P., Desbrun, M., 2011. An optimal
  transport approach to robust reconstruction and simplification of 2{D}
  shapes. In: Computer Graphics Forum. Vol.~30. Wiley Online Library, pp.
  1593--1602.

\bibitem[{Digne et~al.(2014)Digne, Cohen-Steiner, Alliez, De~Goes, and
  Desbrun}]{digne2014feature}
Digne, J., Cohen-Steiner, D., Alliez, P., De~Goes, F., Desbrun, M., 2014.
  Feature-preserving surface reconstruction and simplification from
  defect-laden point sets. Journal of mathematical imaging and vision 48~(2),
  369--382.

\bibitem[{Dorninger and Pfeifer(2008)}]{dorninger2008comprehensive}
Dorninger, P., Pfeifer, N., 2008. A comprehensive automated 3{D} approach for
  building extraction, reconstruction, and regularization from airborne laser
  scanning point clouds. Sensors 8~(11), 7323--7343.

\bibitem[{Guinard and Vallet(2018)}]{guinard2018sensor}
Guinard, S., Vallet, B., 2018. Sensor-topology based simplicial complex
  reconstruction from mobile laser scanning. arXiv preprint arXiv:1802.07487.

\bibitem[{Harary et~al.(2014)Harary, Tal, and Grinspun}]{harary2014context}
Harary, G., Tal, A., Grinspun, E., 2014. Context-based coherent surface
  completion. ACM Transactions on Graphics (TOG) 33~(1), 5.

\bibitem[{Jeong et~al.(2018)Jeong, Yoon, and Park}]{jeong2018multimodal}
Jeong, J., Yoon, T.~S., Park, J.~B., 2018. Multimodal sensor-based semantic 3d
  mapping for a large-scale environment. Expert Systems with Applications 105,
  1--10.

\bibitem[{Kazhdan and Hoppe(2013)}]{kazhdan2013screened}
Kazhdan, M., Hoppe, H., 2013. Screened poisson surface reconstruction. ACM
  Transactions on Graphics (TOG) 32~(3), 29.

\bibitem[{Morsdorf et~al.(2004)Morsdorf, Meier, K{\"o}tz, Itten, Dobbertin, and
  Allg{\"o}wer}]{morsdorf2004lidar}
Morsdorf, F., Meier, E., K{\"o}tz, B., Itten, K.~I., Dobbertin, M.,
  Allg{\"o}wer, B., 2004. Lidar-based geometric reconstruction of boreal type
  forest stands at single tree level for forest and wildland fire management.
  Remote Sensing of Environment 92~(3), 353--362.

\bibitem[{Paparoditis et~al.(2012)Paparoditis, Papelard, Cannelle, Devaux,
  Soheilian, David, and Houzay}]{paparoditis2012stereopolis}
Paparoditis, N., Papelard, J.-P., Cannelle, B., Devaux, A., Soheilian, B.,
  David, N., Houzay, E., 2012. Stereopolis {II}: A multi-purpose and
  multi-sensor 3d mobile mapping system for street visualisation and 3d
  metrology. Revue fran{\c{c}}aise de photogramm{\'e}trie et de
  t{\'e}l{\'e}d{\'e}tection 200~(1), 69--79.

\bibitem[{Popovi{\'c} and Hoppe(1997)}]{popovic1997progressive}
Popovi{\'c}, J., Hoppe, H., 1997. Progressive simplicial complexes. In:
  Proceedings of the 24th annual conference on Computer graphics and
  interactive techniques. ACM Press/Addison-Wesley Publishing Co., pp.
  217--224.

\bibitem[{Pu and Vosselman(2009)}]{pu2009knowledge}
Pu, S., Vosselman, G., 2009. Knowledge based reconstruction of building models
  from terrestrial laser scanning data. ISPRS Journal of Photogrammetry and
  Remote Sensing 64~(6), 575--584.

\bibitem[{Rusu et~al.(2009)Rusu, Blodow, Marton, and Beetz}]{rusu2009close}
Rusu, R.~B., Blodow, N., Marton, Z.~C., Beetz, M., 2009. Close-range scene
  segmentation and reconstruction of 3d point cloud maps for mobile
  manipulation in domestic environments. In: Intelligent Robots and Systems,
  2009. IROS 2009. IEEE/RSJ International Conference on. IEEE, pp. 1--6.

\bibitem[{Rutzinger et~al.(2011)Rutzinger, Pratihast, Oude~Elberink, and
  Vosselman}]{rutzinger2011tree}
Rutzinger, M., Pratihast, A.~K., Oude~Elberink, S.~J., Vosselman, G., 2011.
  Tree modelling from mobile laser scanning data-sets. The Photogrammetric
  Record 26~(135), 361--372.

\bibitem[{Vallet et~al.(2015)Vallet, Br{\'e}dif, Serna, Marcotegui, and
  Paparoditis}]{vallet2015terramobilita}
Vallet, B., Br{\'e}dif, M., Serna, A., Marcotegui, B., Paparoditis, N., 2015.
  Terra{M}obilita/i{Q}mulus urban point cloud analysis benchmark. Computers \&
  Graphics 49, 126--133.

\bibitem[{Xiao et~al.(2013)Xiao, Vallet, and Paparoditis}]{xiao2013change}
Xiao, W., Vallet, B., Paparoditis, N., 2013. Change detection in 3{D} point
  clouds acquired by a mobile mapping system. ISPRS Annals of Photogrammetry,
  Remote Sensing and Spatial Information Sciences 1~(2), 331--336.

\bibitem[{Zhu and Hyyppa(2014)}]{zhu2014use}
Zhu, L., Hyyppa, J., 2014. The use of airborne and mobile laser scanning for
  modeling railway environments in 3d. Remote Sensing 6~(4), 3075--3100.

\bibitem[{Zlot and Bosse(2014)}]{zlot2014efficient}
Zlot, R., Bosse, M., 2014. Efficient large-scale 3{D} mobile mapping and
  surface reconstruction of an underground mine. In: Field and service
  robotics. Springer, pp. 479--493.

\end{thebibliography}

\end{document}